\documentclass{raa}
\usepackage{graphicx,times} 
\usepackage{natbib}
\usepackage{amssymb,amsmath}
\bibpunct{(}{)}{;}{a}{}{,}
\def\degree{\hbox{$^\circ$}}

\usepackage[a4paper=true,pagebackref=true]{hyperref}
\hypersetup{pdftitle = The title of my PDF, pdfauthor = My name, pdfsubject= The subject, pdfkeywords = keyword1 keyword2 keyword3}
\hypersetup{colorlinks = true, linkcolor = green, anchorcolor = red, citecolor = blue, filecolor = red, pagecolor = red, urlcolor = red}
\begin{document}

\title{New continuum and polarization observations of the Cygnus Loop with FAST} \subtitle{I. Data processing and verification}
\volnopage{Vol.0 (20xx) No.0, 000--000}
\setcounter{page}{1}

\author{Xiao-Hui Sun\inst{1}, Mei-Niang Meng\inst{1}, Xu-Yang Gao\inst{2, 3}, Wolfgang Reich\inst{4}, Peng Jiang\inst{2, 3}, Di Li\inst{2,3}, Huirong Yan\inst{5, 6}, Xiang-Hua Li\inst{1}
}
\institute{Department of Astronomy, Yunnan University, Kunming 650500, China; {\it xhsun@ynu.edu.cn}\\
\and
National Astronomical Observatories, Chinese Academy of Sciences, 20A Datun Road, Chaoyang District, Beijing 100101, China; \\
\and
CAS Key Laboratory of FAST, National Astronomical Observatories, Chinese Academy of Sciences, Beijing 100101, China
\and
Max-Planck-Institut f\"ur Radioastronomie, Auf dem H\"ugel 69, 53121 Bonn, Germany\\
\and
Deutsches Elektronen-Synchrotron DESY, Platanenallee 6, D-15738 Zeuthen, Germany\\
\and
Institut f\"ur Physik und Astronomie, Universit\"at Potsdam, Haus 28, Karl-Liebknecht-Str. 24/25, D-14476 Potsdam, Germany\\
\vs\no
{\small Received~~20xx month day; accepted~~20xx~~month day}
}

\abstract{We report on the continuum and polarization observations of the Cygnus Loop supernova remnant (SNR) conducted by the Five-hundred-meter Aperture Spherical radio Telescope (FAST). FAST observations provide high angular resolution and high sensitivity images of the SNR, which will help to disentangle its nature. We obtained Stokes $I$, $Q$ and $U$ maps over the frequency range of 1.03 -- 1.46~GHz split into channels of 7.63~kHz. The original angular resolution is in the range of $\sim3\arcmin$ -- $\sim3\farcm8$, and we combined all the data at a common resolution of $4\arcmin$. The temperature scale of the total intensity and the spectral index from the in-band temperature-temperature plot are consistent with previous observations, which validates the data calibration and map-making procedures. The rms sensitivity for the band-averaged total-intensity map is about 20~mK in brightness temperature, which is at the level of confusion limit. For the first time, we apply rotation measure (RM) synthesis to the Cygnus Loop to obtain the polarization intensity and RM maps. The rms sensitivity for polarization is about 5~mK, far below the total-intensity confusion limit. We also obtained RMs of eight extra-galactic sources, and demonstrate that the wide-band frequency coverage helps to overcome the ambiguity of RM determinations. 
}

\authorrunning{Sun et al.}
\titlerunning{Cygnus Loop with FAST}
\maketitle

\section{Introduction}

Supernova remnants (SNRs) are primary extended radio sources in the Milky Way Galaxy. Radio continuum and polarization observations of SNRs help us to understand their evolution and the magnetic fields in both SNRs and the Galactic interstellar medium.

The Cygnus Loop (G74.0$-$8.5) is a large SNR located at high Galactic latitude, which makes it less obscured by the strong emission from the Galactic plane. This SNR has therefore been extensively observed. In radio band, the Cygnus Loop consists of two bright spherical shell structures centered around similar right ascension (RA). The polarization properties of the two shells are distinct. The northern shell is almost totally depolarized at 1.4~GHz, whereas the depolarization is low toward the southern shell. At higher frequencies such as 5~GHz, both shells show strong polarized emission. However, the magnetic field is nearly tangential to the northern shell but mainly perpendicular to the southern shell. The difference in polarization morphology of the two shells led \citet{uyaniker+02} and \citet{sun+06} to propose that these two shells of the Cygnus Loop are actually two individual SNRs with mutual interactions. High angular resolution and high sensitivity radio continuum and polarization images of the Cygnus Loop, and the maps of spectral indices and rotation measures (RMs) derived from these images will help to differentiate the two shells and therefore shed light on the nature of the SNR. 

A large field of view is required for mapping extended objects such as the Cygnus Loop with an angular size of about $4\degr\times4\degr$. Multi-beam receivers have been widely used 
to increase the field of view and hence the survey ``speed". The L-band 13-beam receiver at the Parkes 64-m telescope~\citep{staveley-smith+96} has made huge impacts by delivering the pulsar survey that discovered more than half of the known pulsars~\citep{camilo+00}, the southern-sky Galactic HI survey~\citep{mcclure-griffiths+09} and the southern-sky extra-galactic HI surveys~\citep{staveley-smith+00}. The L-band 7-beam receiver at the Arecibo 305-m telescope has also been successful in conducting the HI survey~\citep{giovanelli+05, peek+18} and the continuum survey~\citep{taylor+salter+10}. Also at the Effelsberg 100-m telescope a L-band 7-beam prime-focus receiver is extensively used for pulsar observations \citep{Barr+13}. The Five-hundred-meter Aperture Spherical radio Telescope (FAST, \citealp{nan+11}) is equipped with a 19-beam L-band receiving system, which is the largest multi-beam receiver of its kind.

FAST is the largest single-dish telescope in the world and has been conducting scientific observations since 2020. \citet{jiang+19} and \citet{jiang+20} presented a detailed description of the technical specifications and performance of FAST. Its large collecting area guarantees high angular resolution and sensitivity. The 19-beam L-band receiving system at FAST largely increases the survey ``speed" so that extended objects such as the Cygnus Loop can be imaged. The Commensal Radio Astronomy FAST Survey (CRAFTS)\footnote{https://crafts.bao.ac.cn/} is one of the FAST legacy projects to simultaneously deliver all-sky surveys of pulsars, Galactic and extra-galactic HI, and continuum and polarization~\citep{li+18}. The summary of the scientific achievements and perspective of FAST was presented by \citet{qian+20}.

In two papers, we present new continuum and polarization observations of the Cygnus Loop conducted by FAST. In this paper, we focus on data processing and verification, aiming to demonstrate the imaging capability of FAST. In Paper II, we will present a detailed analysis of the nature of the Cygnus Loop. The paper is organized as follows. We describe the observations and data processing in Sect.~\ref{sect:obs}, present results and verification in Sect.~\ref{sect:results}, and summarize the results in Sect.~\ref{sect:conclusion}.

\section{Observations and data processing}
\label{sect:obs}

The observations were conducted during the commissioning of FAST as a shared-risk project (code: 2019a-125-C). The drifting-scan mode~\citep{jiang+20} was used to map the Cygnus Loop. In this mode, the L-band 19-beam receiver at the focus of FAST is rotated by 23$\fdg4$, so that a scan covers a width of about 20$\arcmin$ with full Nyquist sampling. The scanning velocity is about $15\arcmin/\cos(\delta)$ min$^{-1}$, where $\delta$ is the declination. We mapped the Cygnus Loop in RA direction only. In total, 15 drifting scans were made in four time slots in 2019: April 29, May 02, May 04, and May 12. With the mode of ``on-the-fly" mapping, we also made maps of the compact radio sources 3C~138 and 3C~48 with each individual beam for calibration. The total observation time was about 12 hours. 

The 19-beam receiver was installed at the focal feed cabin supported by six cables~\citep{jiang+19}. By tuning positions of the feed cabin and the active surface reflectors, a paraboloid of 300-m diameter is formed pointing to the target direction. The signal is converted to an optical signal in the cabin, transferred via an optical fiber to the control room, converted back to an electric signal and digitized. The entire FAST backend contains 12 ROACH-2 backends that enable simultaneous data recording of spectral lines, wide band continuum and polarization, and pulsars~\citep{jiang+19}.

The data dumping rate is about 1~s. A linearly polarized reference signal produced by a diode is injected into the system every other 1~s, which is used to establish the temperature scale and correct the system drifting during scans. The temperature for the reference signal has two levels: $\sim$12.5~K and $\sim$1.1~K, for observations of strong sources such as calibrators and program sources such as the Cygnus Loop, respectively. The frequency range for the receiving system is 1.0 -- 1.5~GHz split into 65536 channels. Each individual channel has a width of about 7.63~kHz. The output from the backend contains two channels for total intensity $I_1$ and $I_2$, and two channels for Stokes $U$ and $V$.   
We conducted radio frequency interference (RFI) flagging based on one of the observations of 3C~138. For each frequency channel, we calculated the root mean square (rms). A median filter was then applied to all the rms to obtain the median at each frequency, and the median absolute deviation (MAD)\footnote{MAD is the median of the absolute deviation from the median. Both mean and MAD are more robust than mean and standard deviation for data with outliers. This is the case here as RFIs still remain in the data.}. The frequency channel with rms deviating from the median by larger than $5\times$MAD was flagged. This process was repeated for $I_1$, $I_2$, $U$ and $V$. We applied RFI flagging to all the other observations. Additionally, about 4000 frequency channels at both ends of the 500~MHz band were also removed because of low gain values. We also flagged the channels with contributions from the HI line emission. We were left with about 29\,500 channels, which is sufficient for the Cygnus Loop since it is very bright. A more thorough RFI mitigation procedure is being developed to preserve more data.  

We processed the data from the calibrators and obtained maps per frequency channel per beam. Based on these maps, we derived beam widths, gain values and polarization properties of the system.

\subsection{Calibration and Map-making}

 For each scan, we compared the intensity levels in arbitrary units from the receiving system with and without the injected reference signal to obtain the intensity levels in K~$T_{\rm a}$ (antenna temperature). The FAST 19-beam receiver uses linear feeds for polarization observations. We thus derived the Stokes parameters as $I=I_1+I_2$, $Q=I_1-I_2$, $U$, and $V$. Note that $Q$ is the difference of the two total-intensity channels and may suffer from fluctuating gain ratios of the channels, which makes it challenging to calibrate $Q$. In contrast, circular feeds yield both $Q$ and $U$ as correlations of the left- and right-handed circular components and therefore more ideal for linear-polarization observations. 
 
 The mismatch between the amplitudes of the gains of the two linear feeds causes leakage between $I$ and $Q$, which can be written for the injected reference signal as, 
 \begin{align}
 I_{\rm obs}^{\rm ref} &= I_{\rm true}^{\rm ref} + f Q_{\rm true}^{\rm ref}\\
 Q_{\rm obs}^{\rm ref} &= Q_{\rm true}^{\rm ref} + f I_{\rm true}^{\rm ref},
 \end{align}
 where ``ref" stands for the injected reference signal, ``obs" stands for the observed values, ``true" stands for true values, and $f$ is the leakage factor between $I$ and $Q$. The mismatch between the phases of the gains of the two linear feeds, meaning the two feeds are not orthogonal, can be characterized with the instrumental polarization angle $\chi$ as
 \begin{align}
U_{\rm obs}^{\rm ref} & = \sqrt{U_{\rm true}^{\rm ref\,2}+V_{\rm true}^{\rm ref\,2}} \cos(2\chi) \\
V_{\rm obs}^{\rm ref} & = \sqrt{U_{\rm true}^{\rm ref\,2}+V_{\rm true}^{\rm ref\,2}} \sin(2\chi).
 \end{align}
 Ideally, both $f$ and $\chi$ should be zero. Since the injected reference signal is linearly polarized with equal intensities in $I_1$ and $I_2$, namely, $Q_{\rm true}^{\rm ref}=V_{\rm true}^{\rm ref}=0$, and $I_{\rm true}^{\rm ref}=U_{\rm true}^{\rm ref}$, we can obtain $f$ and $\chi$ as,
 \begin{equation}
 f=\frac{Q_{\rm obs}^{\rm ref}}{I_{\rm obs}^{\rm ref}},\,\,\,\,\chi=\frac{1}{2}\arctan\frac{V_{\rm obs}^{\rm ref}}{U_{\rm obs}^{\rm ref}}. 
 \end{equation}
 The absolute value of $f$ varies with beams and time, and is generally smaller than about 20\%. The instrumental polarization angle $\chi$ is beam- and frequency-dependent in the range of $-90\degree<\chi<90\degree$. 
 
Applying the leakage corrections to the observed sky signal of $I$ and $Q$, 
  \begin{align}
 I_{\rm obs}^{\rm sky} &= I_{\rm true}^{\rm sky} + f Q_{\rm true}^{\rm sky}\\
 Q_{\rm obs}^{\rm sky} &= Q_{\rm true}^{\rm sky} + f I_{\rm true}^{\rm sky},
 \end{align}
 and the de-rotation of the instrumental polarization angle for $U$ and $V$, we can obtain the follows: 
\begin{align}
I_{\rm true}^{\rm sky} & = \frac{I_{\rm obs}^{\rm sky}-f Q_{\rm obs}^{\rm sky}}{1-f^2} \\
Q_{\rm true}^{\rm sky} & = \frac{Q_{\rm obs}^{\rm sky}-f I_{\rm obs}^{\rm sky}}{1-f^2} \\
U_{\rm true}^{\rm sky} & = U_{\rm obs}^{\rm sky} \cos(2\chi) + V_{\rm obs}^{\rm sky} \sin(2\chi) \\
V_{\rm true}^{\rm sky} & = -U_{\rm obs}^{\rm sky} \sin(2\chi) + V_{\rm obs}^{\rm sky} \cos(2\chi)
\end{align}
Here, the superscript ``sky" stands for the sky signal, and the subscripts ``obs" and ``true" represent the observed and corrected quantities, respectively. A baseline by linearly fitting the two ends of a scan was subsequently subtracted to remove unrelated large scale emission and the system temperature offset. We then combined all the baseline-corrected scans by interpolation to obtain maps of $I$, $Q$, $U$ and $V$ for each beam and frequency in antenna temperature $T_a$.

\begin{figure}[!htbp]
    \centering
    \includegraphics[width=0.95\textwidth]{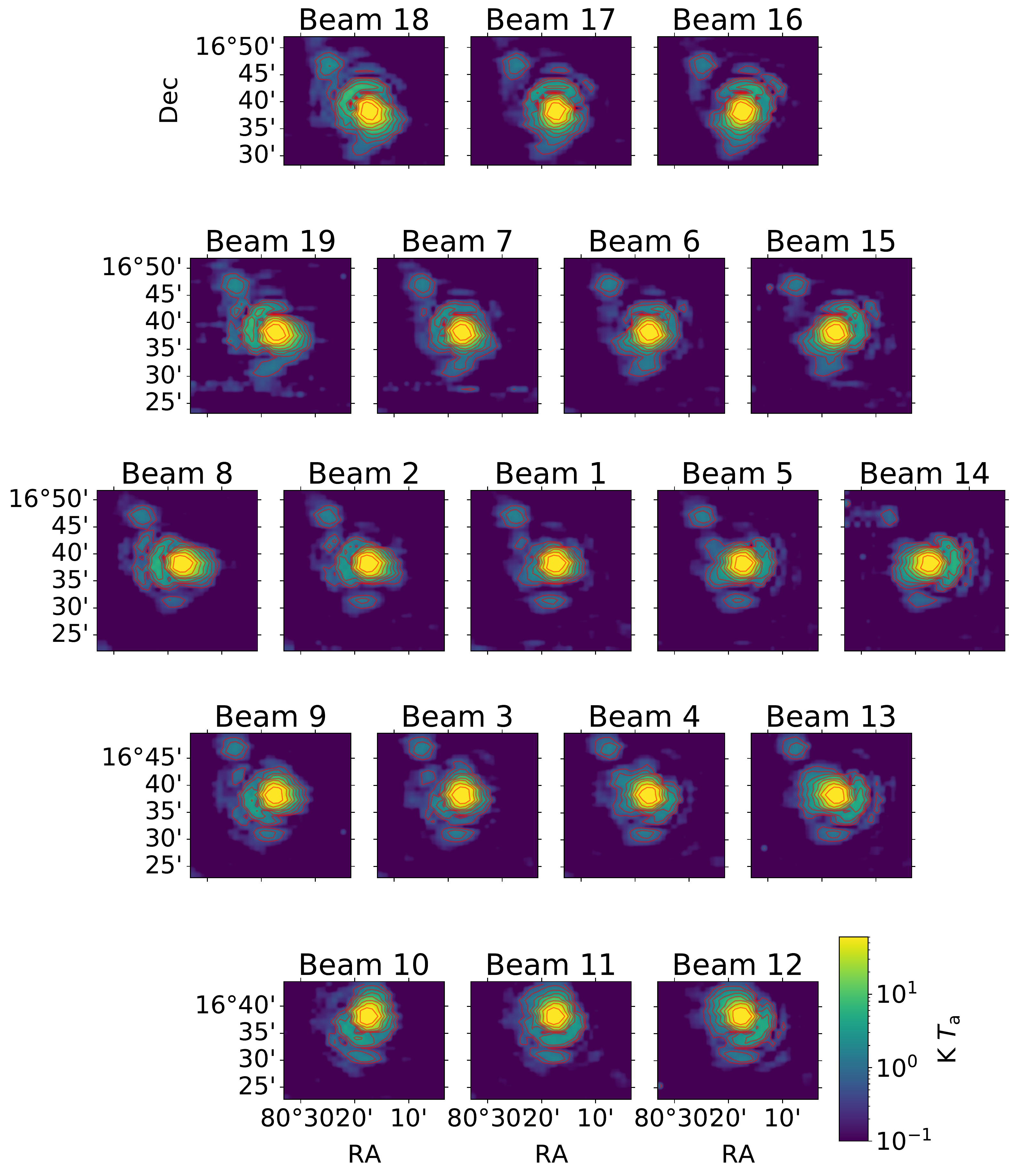}
    \caption{Total-intensity images of 3C~138 for the 19 beams at 1401.912~MHz in logarithmic color scale. The maps are in unit of antenna temperature (~K~$T_{\rm a}$). The contours run from $-3$~dB to $-24$~dB in steps of $-3$~dB.}
    \label{fig:beam}
\end{figure}

We show the total-intensity images of 3C~138 for the 19 beams at the frequency channel of 1401.912~MHz in Fig.~\ref{fig:beam} , illustrating the beam patterns. In order to increase the signal to noise ratio, we averaged 400 frequency channel maps around this frequency. The images are arranged according to the layout of the beams~\citep[][their Fig. 22]{jiang+20}, and Beam 1 is the central beam on focus. If a beam is off focus along a certain direction, an asymmetric coma lobe is expected in the opposite direction, which can be seen for all the other beams. The levels of these coma lobes are roughly in the range of 0.4\%-6\% of the peak values. Although the bandwidth of about 7.63~kHz for each frequency channel is narrow, 3C~138 is very bright and the signal-to-noise ratio is thus sufficiently high allowing us to make channel-by-channel analyses. 

\subsection{Beam, gain and brightness temperature}
\label{sect:gain}
\begin{figure}[!htbp]
    \centering
    \includegraphics[width=0.8\textwidth]{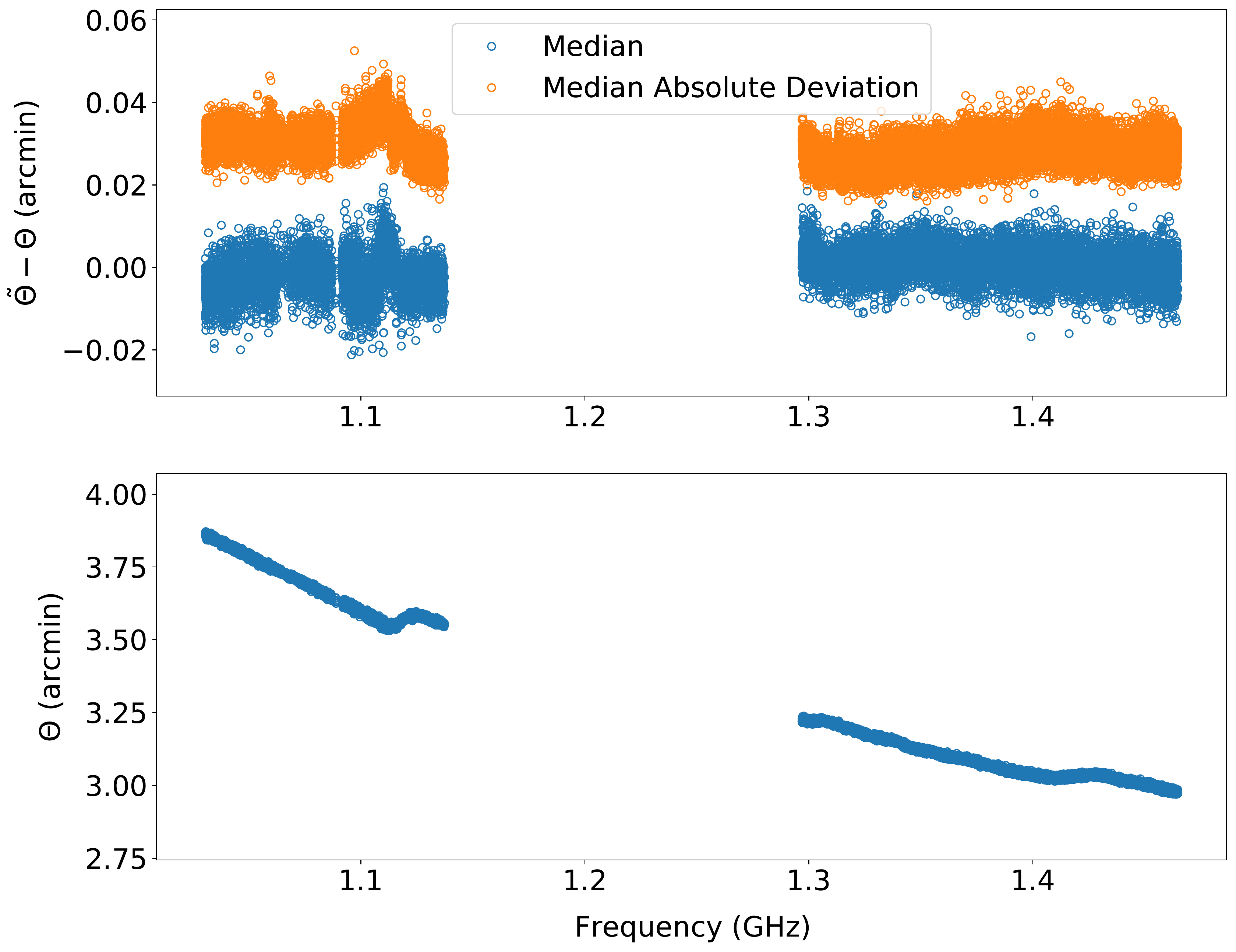}
    \caption{Beam width versus frequency. Bottom: the median beam width ($\Theta$) at each frequency; Top: the difference between the beam width ($\tilde{\Theta}$) of different beams at different time and the median.}
    \label{fig:beam_width}
\end{figure}

We mapped 3C~138 for three times with each of the 19 beams, so that there are 57 maps at each frequency channel. We made 2D elliptical Gaussian fittings to the total-intensity maps for each beam at each frequency channel to obtain the peak temperature, and the full width half magnitude as the major and minor beam width. The difference between the major and minor beam width is very small. We used their average as the beam width $\tilde{\Theta}$ hereafter. To obtain Stokes $Q$ and $U$, we used the values at the positions of total-intensity peaks. The median value $\Theta$ of the beam widths of all the 57 maps can also be obtained. The results are shown in Fig.~\ref{fig:beam_width} for all the frequency channels. The difference between $\tilde{\Theta}$ and $\Theta$ is negligible with a median of about 0 and a median absolute deviation of about 0.03 arcmin. This means that the beam width is nearly the same for all the beams and does not vary over the time of the whole observations spanning about two weeks. We used $\Theta$ for the following analyses. 

As can be seen from Fig.~\ref{fig:beam_width}, $\Theta$ is about $3\farcm8$ at the low-frequency end and about $3\arcmin$ at the high-frequency end. The relation between $\Theta$ and frequency is not continuous and has two jumps at $\sim$1.12~GHz and $\sim$1.42~GHz. In general, $\Theta$ is not proportional to the reciprocal of frequency, as has also been shown by \citet{jiang+20}.

\begin{figure}[!htbp]
    \centering
    \includegraphics[width=0.8\textwidth]{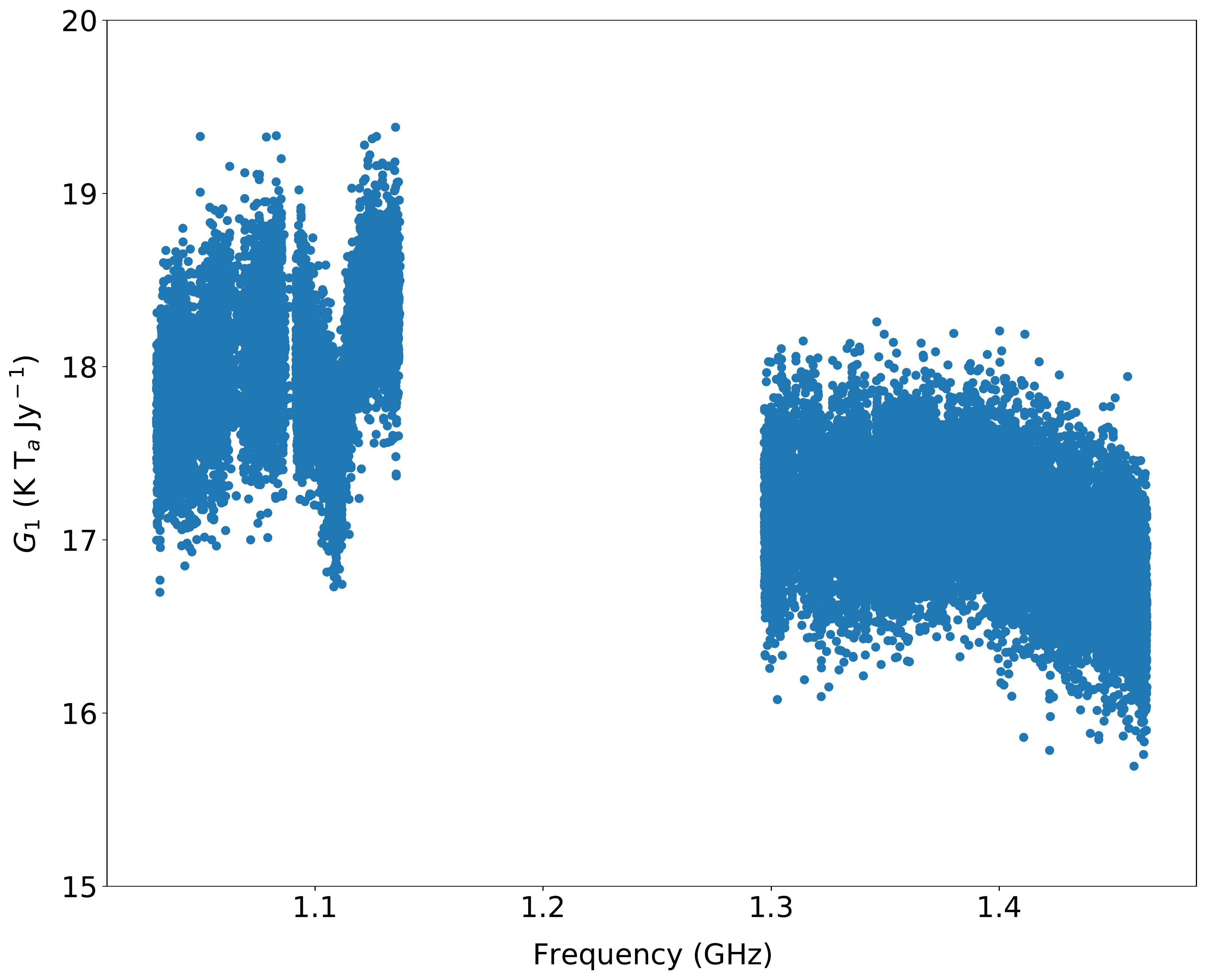}
    \caption{Gain of the central beam, $G_1$, versus frequency.}
    \label{fig:gain1}
\end{figure}

The gain ($G$) of FAST can be measured from the observations of calibrators as $G=T/S$, where $T$ is the total intensity $I$ of the source in antenna temperature derived from the Gaussian fittings, and $S$ is the flux density of the source. We used the model by \citet{perley+butler+17} to calculate the flux densities of 3C~138 at given frequencies. Similar to the beam width, the variation of $G$ over time is small, and we used the median values of the observations at the three time slots, and show the gain of the central beam, $G_1$, in Fig.~\ref{fig:gain1}. The gain is about 18~K~Jy$^{-1}$ at the low frequency end and about 17~K~Jy$^{-1}$ at the high frequency end. Note that the gains obtained by \citet{jiang+20} are smaller than our values, and one of the reasons could be the variation of the temperature of the injected noise. The effective illumination diameter is 300 m for FAST, corresponding to a perfect gain of 25.6~K~Jy~$^{-1}$~\citep{jiang+20}. In our case, this means an aperture efficiency of about 70\% at the low-frequency end, and about 66\% at the high-frequency end.  

\begin{figure}[!htbp]
    \centering
    \includegraphics[width=0.8\textwidth]{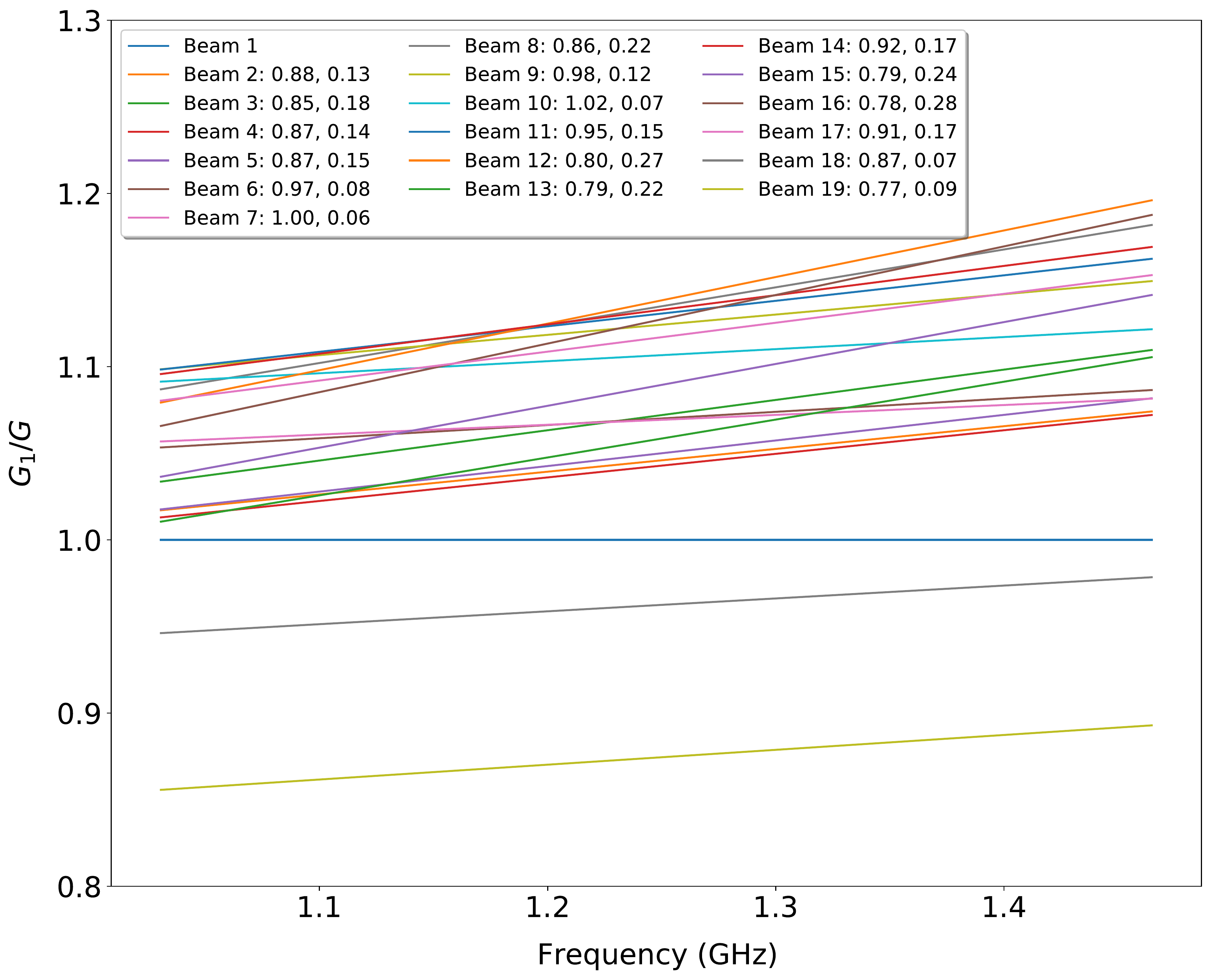}
    \caption{Ratio of gains between the outer beams and the central beam (Beam 1) fitted to a straight line. The intercepts and slopes for the fittings are shown in the legend.}
    \label{fig:gain_ratio}
\end{figure}

For the observations of the Cygnus Loop, one drifting scan contains 19 sub-scans from each of the beams. The variations of $G$ among different beams produce stripes along the scan direction, which is the primary reason for scanning effects besides baseline uncertainties. The variations of $G$ against frequency are also beam-dependent, which will complicate the calculation of the spectrum of the Cygnus Loop. To account for the varying $G$ versus beams and frequencies, we normalized the gains with respect to the gain of the central beam (Beam 1). The ratio $G_1/G$ versus frequency for the other 18 beams can be fit to a straight line, as shown in Fig.~\ref{fig:gain_ratio}. The intercepts and slopes of the linear fittings are also displayed in Fig.~\ref{fig:gain_ratio}. The factor $G_1/G$ from the fittings was applied to the sub-scans of corresponding beams for map-making of the Cygnus Loop, which largely suppressed the stripes from the scanning effects.

The intensity often is represented in main-beam brightness temperature $T_{\rm b}$ instead of antenna temperature $T_{\rm a}$, since the main beam can be directly measured. The relation between these two temperatures can be written as 
\begin{equation}
T_{\rm b}=\frac{\Omega_{\rm a}}{\Omega_{\rm b}}T_{\rm a}=\frac{T_{\rm a}}{\eta_{\rm b}},
\end{equation}
where $\Omega_{\rm a}$ is the antenna solid angle, $\Omega_{\rm b}=1.13\Theta^2$ is the main beam solid angle, and $\eta_{\rm b}=\Omega_{\rm b}/\Omega_{\rm a}$ is the main beam efficiency. From observations of the calibrators, we can derive the antenna solid angle as 
\begin{equation}
\Omega_{\rm a}=\frac{\lambda^2 S}{2kT_{\rm a}}, 
\end{equation}
where $\lambda$ is the observing wavelength, and $k$ is the Boltzmann's constant. We found that $\eta_{\rm b}$ is about 80\%-85\% at low-frequency end, and about 90\%-95\% at high-frequency end. Note that the temperature of the injected noise used in this paper is larger than the recent measurements by about 4\%. Taking this into account, the main beam efficiency at the high frequency end would be about 90\%. Since all the gains have been normalized to the central beam, the brightness temperature can be obtained by dividing the main-beam efficiency of the central beam. The fluctuation of the main-beam efficiency relative to the average is about 2\%, corresponding to an uncertainty of about 2\% for the brightness temperature. 

The conversion factor, $G_{{\rm Jy}\rightarrow T_{\rm b}}$, from flux density to brightness temperature at an angular resolution of $\Theta_o$ can be written as, 
\begin{equation}\label{eq:jy_k}
    G_{{\rm Jy}\rightarrow T_{\rm b}} = \frac{G}{\eta_{\rm b}} \frac{\Theta^2}{\Theta_o^2}.
\end{equation}

\subsection{Polarization}
\begin{figure}[!htbp]
    \centering
    \includegraphics[width=0.95\textwidth]{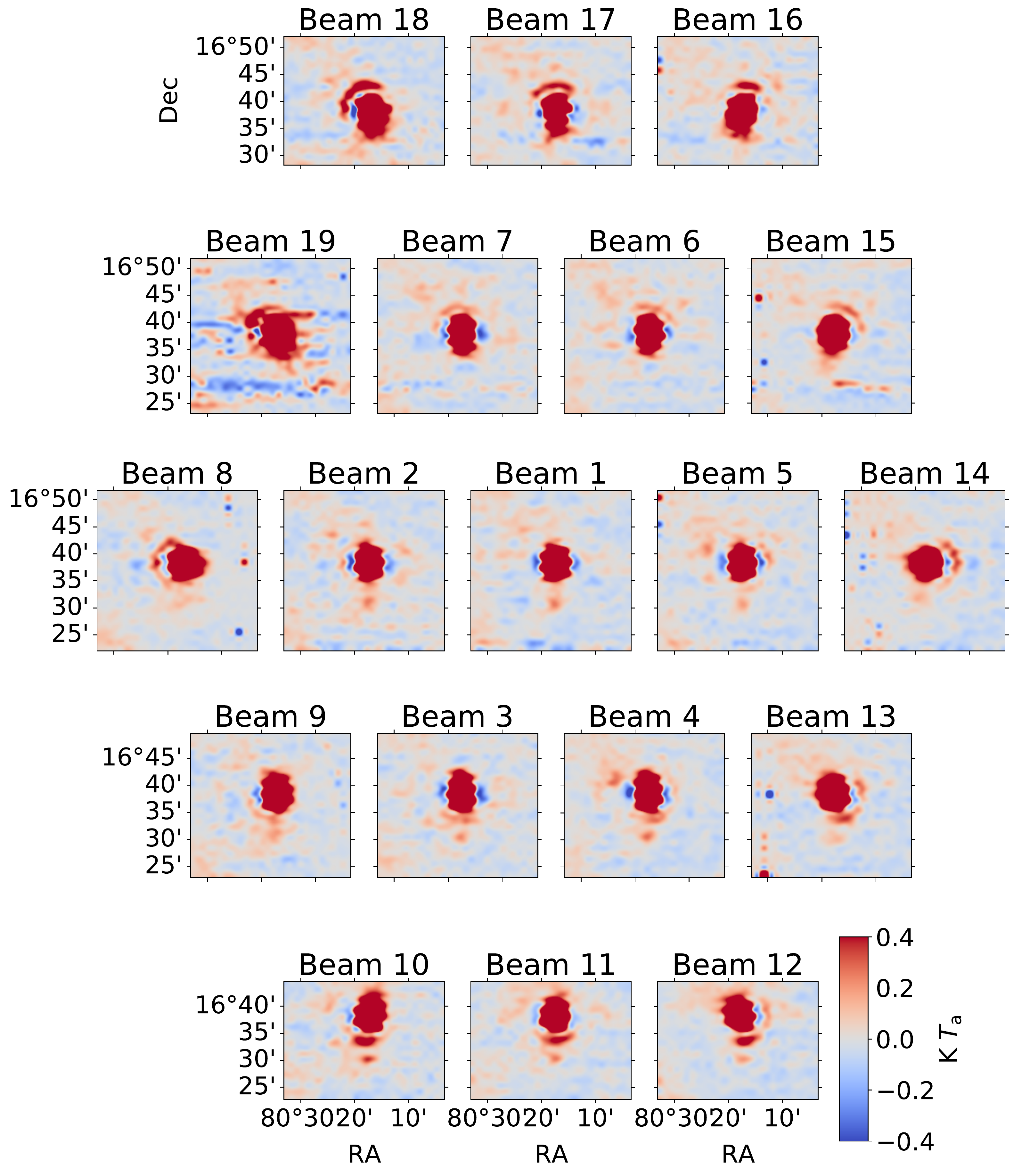}
    \caption{Stokes $Q$ images of 3C~138 for the 19 beams at 1401.912~MHz.}
    \label{fig:beam_Q}
\end{figure}

\begin{figure}[!htbp]
    \centering
    \includegraphics[width=0.95\textwidth]{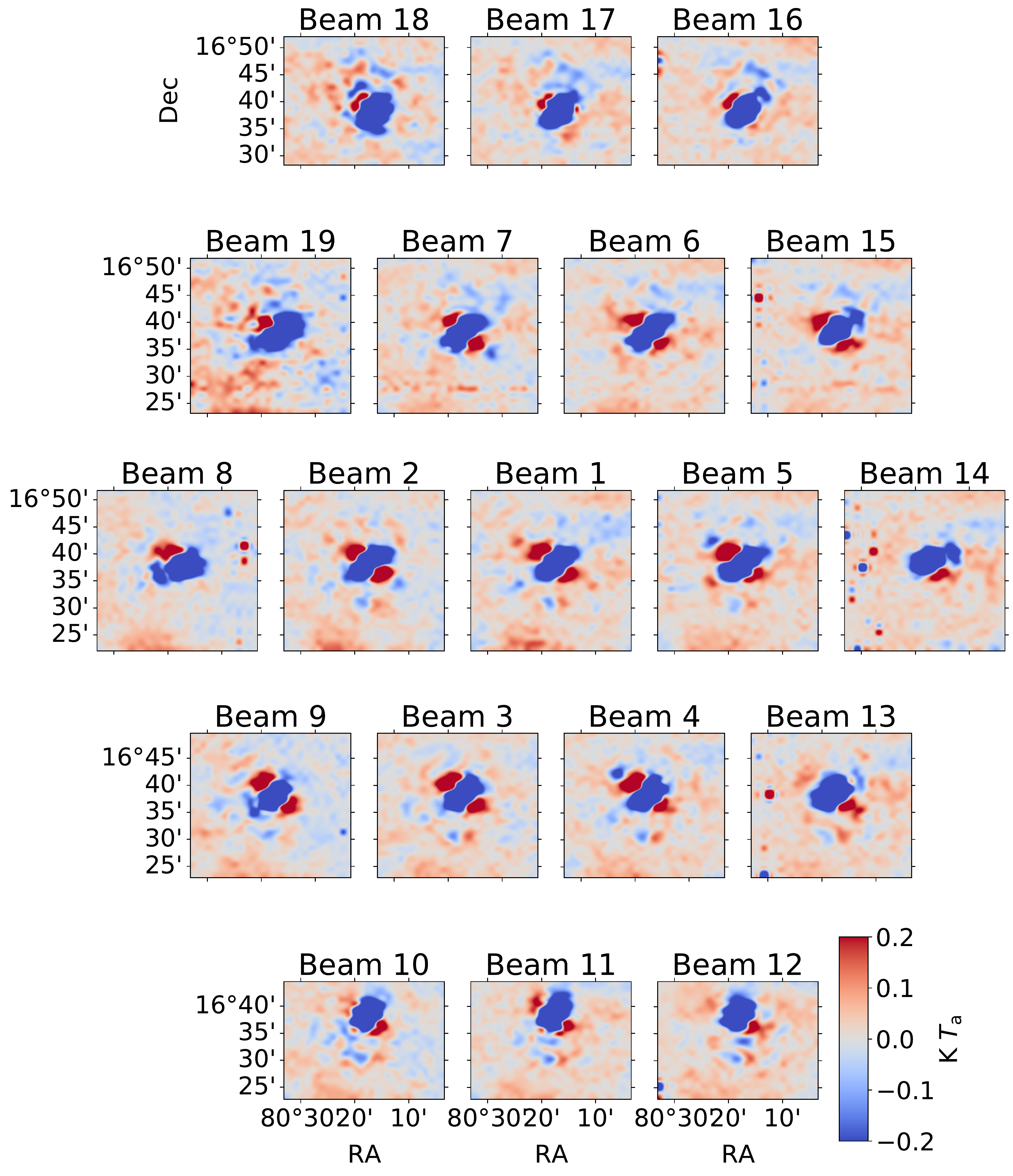}
    \caption{Stokes $U$ images of 3C~138 for the 19 beams at 1401.912~MHz.}
    \label{fig:beam_U}
\end{figure}

\begin{figure}[!htbp]
    \centering
    \includegraphics[width=0.95\textwidth]{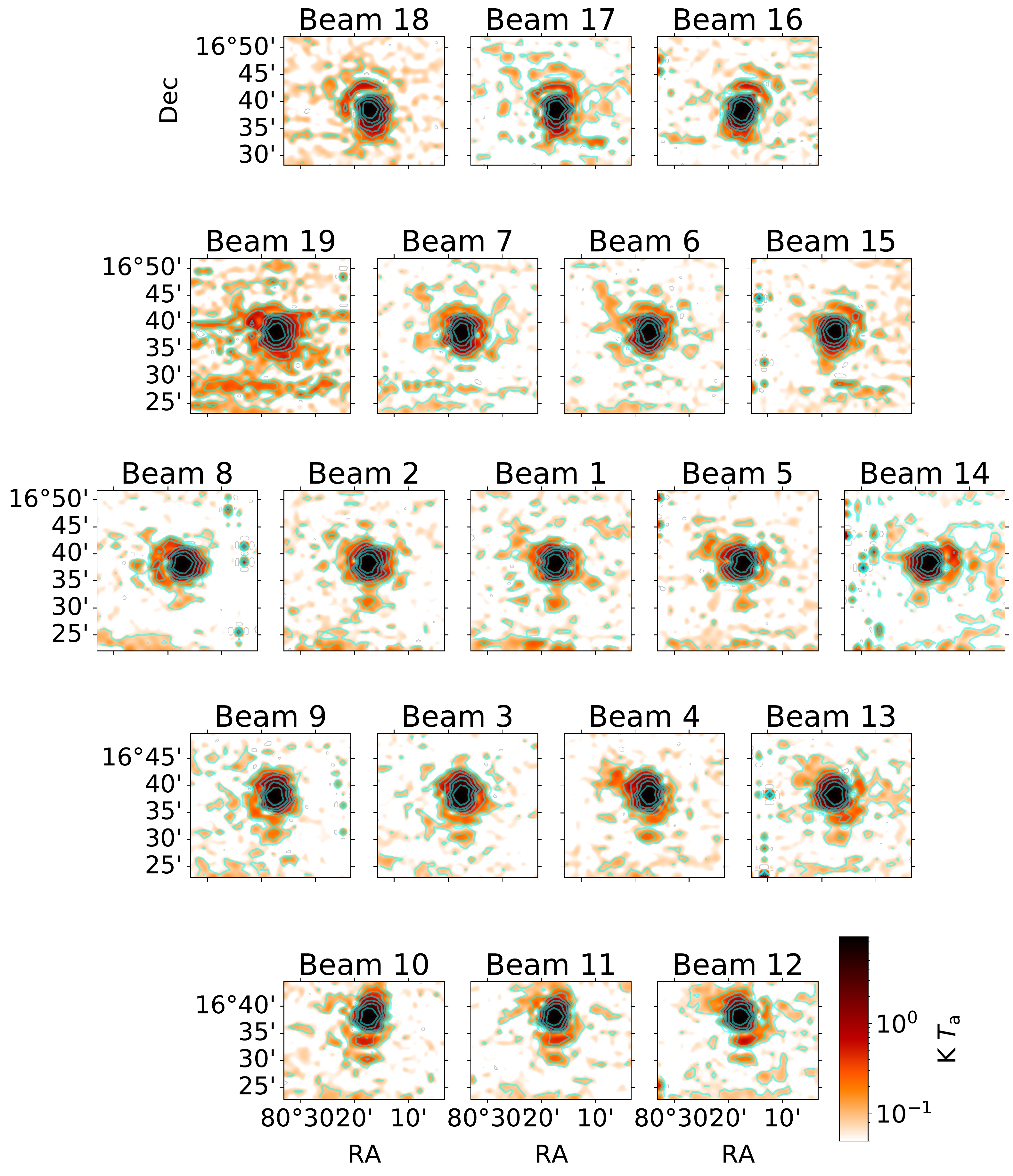}
    \caption{Polarized intensity images of 3C~138 for the 19 beams at 1401.912~MHz in logarithmic color scale. The contours run from $-3$~dB to $-21$~dB in a step of $-3$~dB in respect
    to the polarized intensity maximums.}
    \label{fig:beam_PI}
\end{figure}

The polarized intensity ($P$) can be calculated as $P=\sqrt{Q^2+U^2}$, and the degree of polarization is then $\Pi=P/I$. The polarization angle $\psi$ can be derived as $\psi=\frac{1}{2}\arctan\frac{U}{Q}$. When the linearly polarized wave propagates in the magnetized interstellar medium, the polarization angle is rotated as $\Delta\psi\propto {\rm RM}\,\lambda^2$. Here RM is the rotation measure which is the integration of thermal electron density weighted by the line-of-sight magnetic field. The $Q$, $U$, and $P$ images of 3C~138 for each individual beam are shown in Figs.~\ref{fig:beam_Q}, \ref{fig:beam_U}, and \ref{fig:beam_PI}. As can be seen from these figures, $Q$ is peaked at the center, $U$ manifests butterfly pattern, and $P$ is also peaked at the center. Coma lobes can be clearly seen in $P$ with levels up to about 6\% of the peaks of the polarized intensity. 

\begin{figure}[!htbp]
    \centering
    \includegraphics[width=0.8\textwidth]{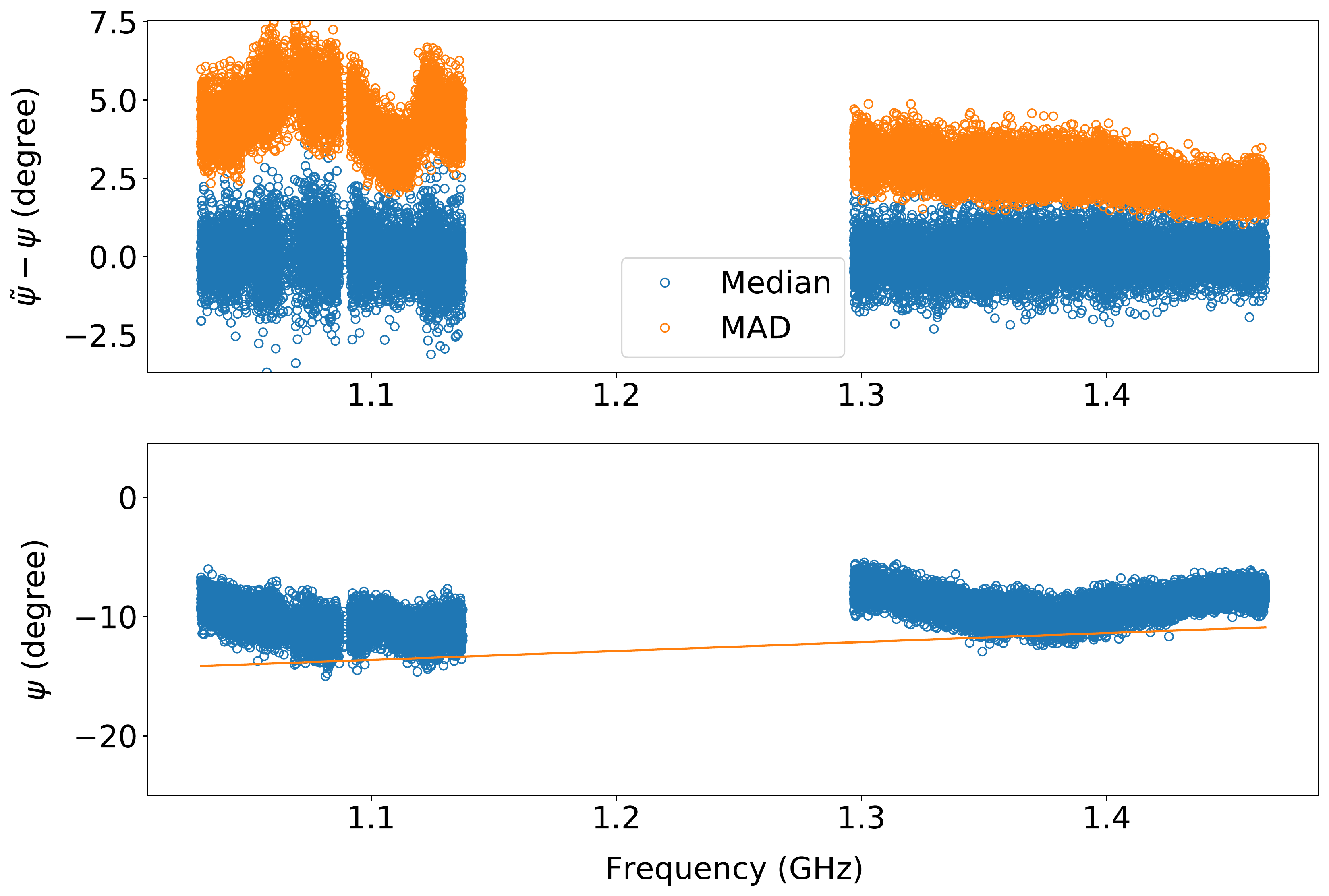}
    \caption{Bottom panel: the median values of all the polarization angles from different time and beams. The model from \citet{perley+butler+13} is represented by the red line. Top panel: median and median absolute deviation of the difference between the polarization angles at different time and beams and the median shown in the bottom panel.}
    \label{fig:pa_diff_ave}
\end{figure}

\begin{figure}[!htbp]
    \centering
    \includegraphics[width=0.8\textwidth]{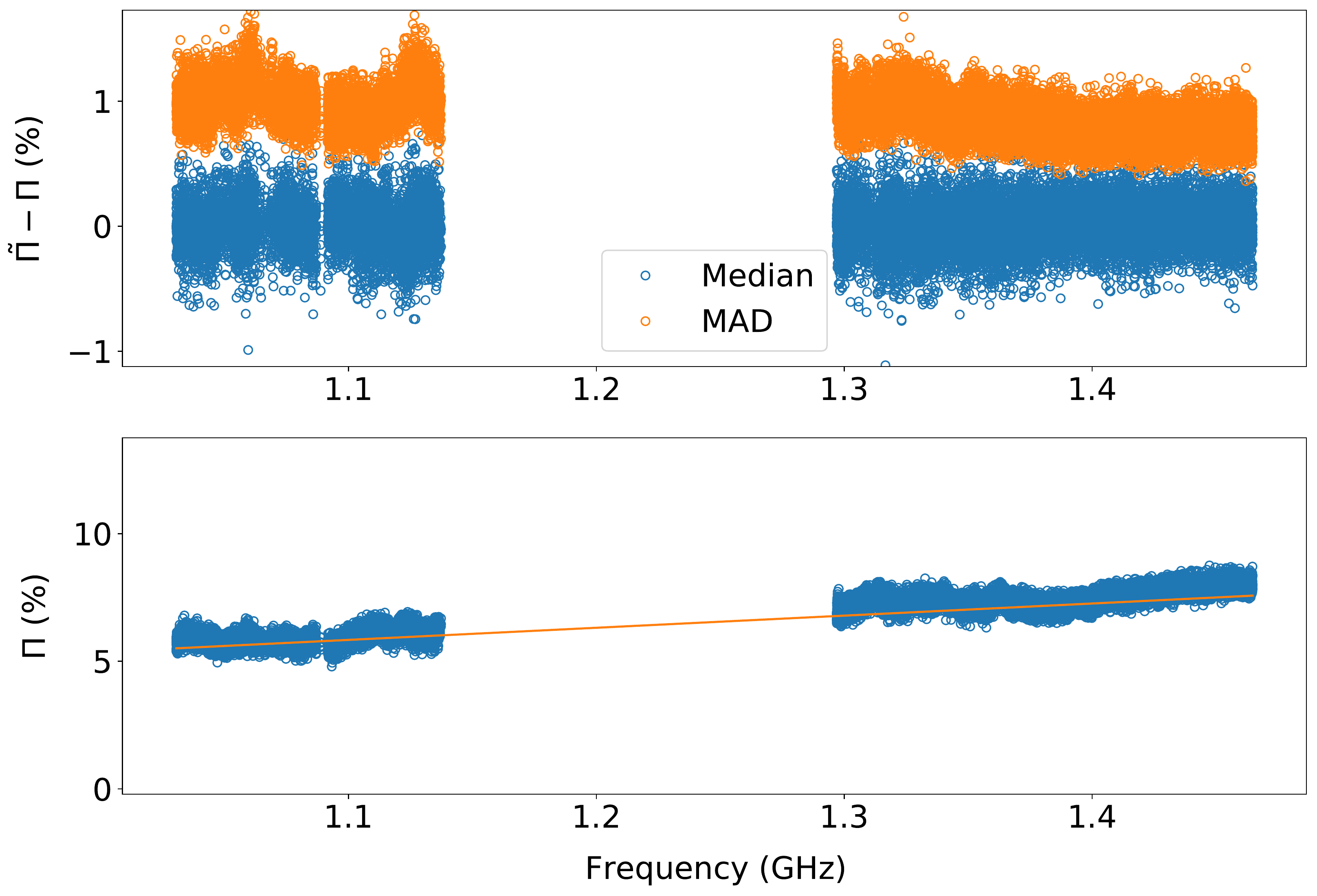}
    \caption{The same as Fig.~\ref{fig:pa_diff_ave} but for degree of polarization.}
    \label{fig:pc_diff_ave}
\end{figure}

For each of the three observations of 3C~138 with each of the 19 beams, we can obtain a polarization angle $\tilde{\psi}$ and a fractional polarization $\tilde{\Pi}$. For each frequency, we estimated the median $\psi$ and $\Pi$ from the 57 values of $\tilde{\psi}$ and $\tilde{\Pi}$. For the polarization angle, the difference between the individual value and the median is around 0 with a median absolute deviation of about $5\degree$ at the low-frequency end and about $3\degree$ at the high-frequency end (Fig.~\ref{fig:pa_diff_ave}). For the degree of polarization, the difference between the individual value and the median value has a mean of about 0 and a median absolute deviation of about 1\% (Fig.~\ref{fig:pc_diff_ave}). The variations of the polarization angle and degree of polarization versus beams and time are also small, similar to the beam width. We will use the median values hereafter. 

For 3C~138, the polarization angle is $-14\degree$ at 1.05~GHz and $-11\degree$ at 1.45~GHz, and the degree of polarization is 5.6\% at 1.05~GHz and 7.5\% at 1.45~GHz from the measurements by \citet{perley+butler+13}. We made a linear interpolation to obtain polarization angle and degree of polarization at other frequencies as the model values, which are represented by red lines in Figs.~\ref{fig:pa_diff_ave} and \ref{fig:pc_diff_ave}. For the polarization angle, the difference between the median $\psi$ and the model values has a median of about 3$\degree$ and a scattering of about 1$\degree$. The difference of $3\degree$ corresponds to an RM of about 1~rad~m$^{-2}$ at the frequency range of 1.05 -- 1.45 GHz, which is smaller than the RM fluctuation of about 5~rad~m$^{-2}$ inside the Cygnus Loop. We therefore did not make further corrections for the polarization angle. For the {degree of polarization}, the observed values are well consistent with the model values. 

For 3C~48, according to the measurements by \citet{perley+butler+13}, its degree of polarization is too low, which is about 0.3\% at 1.05~GHz and 0.5\% at 1.45~GHz, and the polarization angle has a dramatic change from 25$\degree$ at 1.05~GHz to 140$\degree$ at 1.45~GHz. From our observations, the degree of polarization is indeed very low, which makes it unsuitable for polarization calibration. We therefore did not use 3C~48 for calibration. 

\section{Results and verification}
\label{sect:results}
The processing of the Cygnus Loop was very similar to that of the calibrators except for correcting the variation of the gains versus beams and frequencies. We obtained $I$, $Q$, and $U$ maps for the Cygnus Loop at each individual frequency channel, and the scale of the maps was set to main beam brightness temperature $T_{\rm b}$. We smoothed all the maps to a common angular resolution of $4\arcmin$ to derive the frequency cubes of $I$, $Q$, and $U$. 

\subsection{Total intensity}
\begin{figure}
    \centering
    \includegraphics[width=0.75\textwidth]{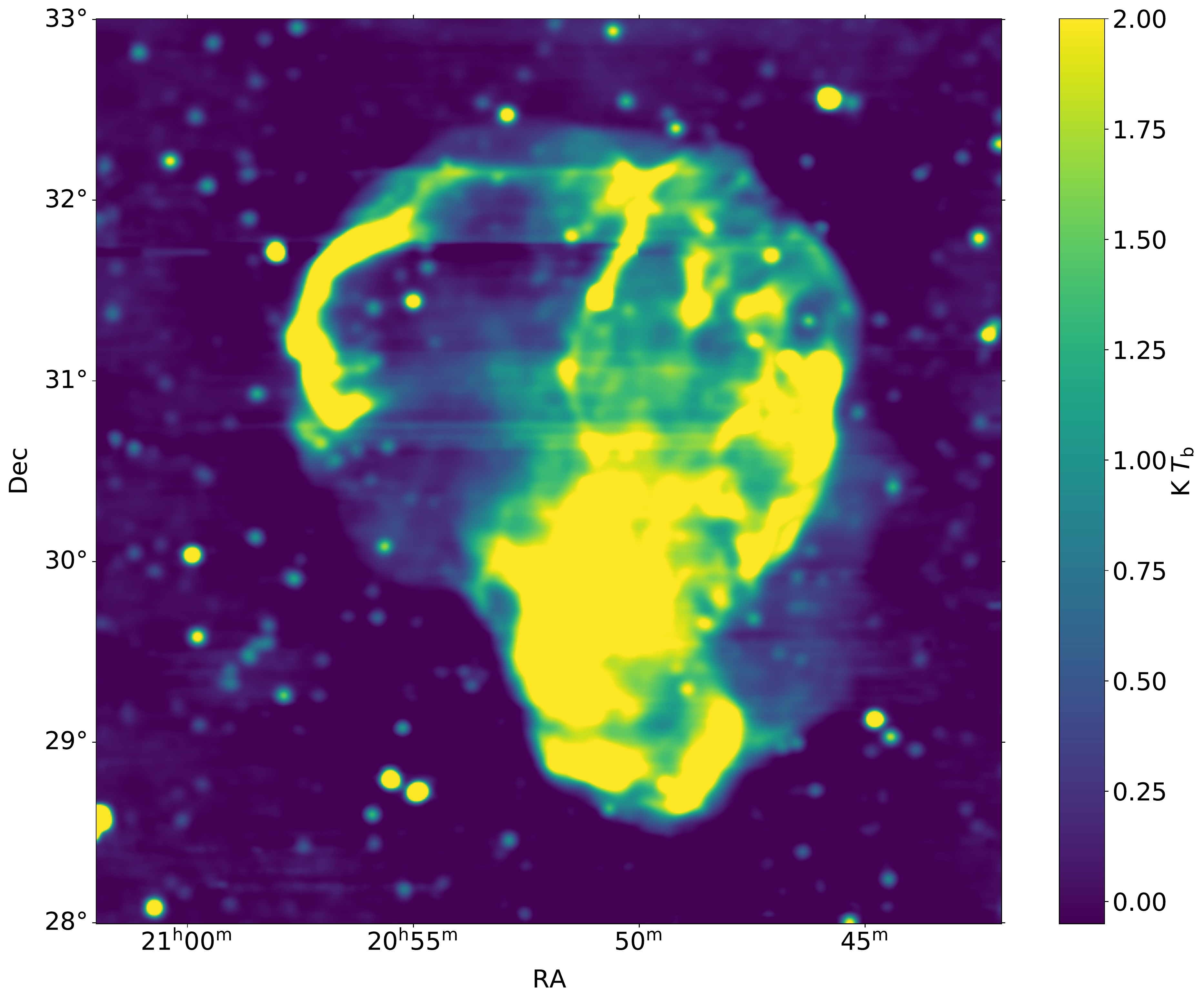}
    \includegraphics[width=0.75\textwidth]{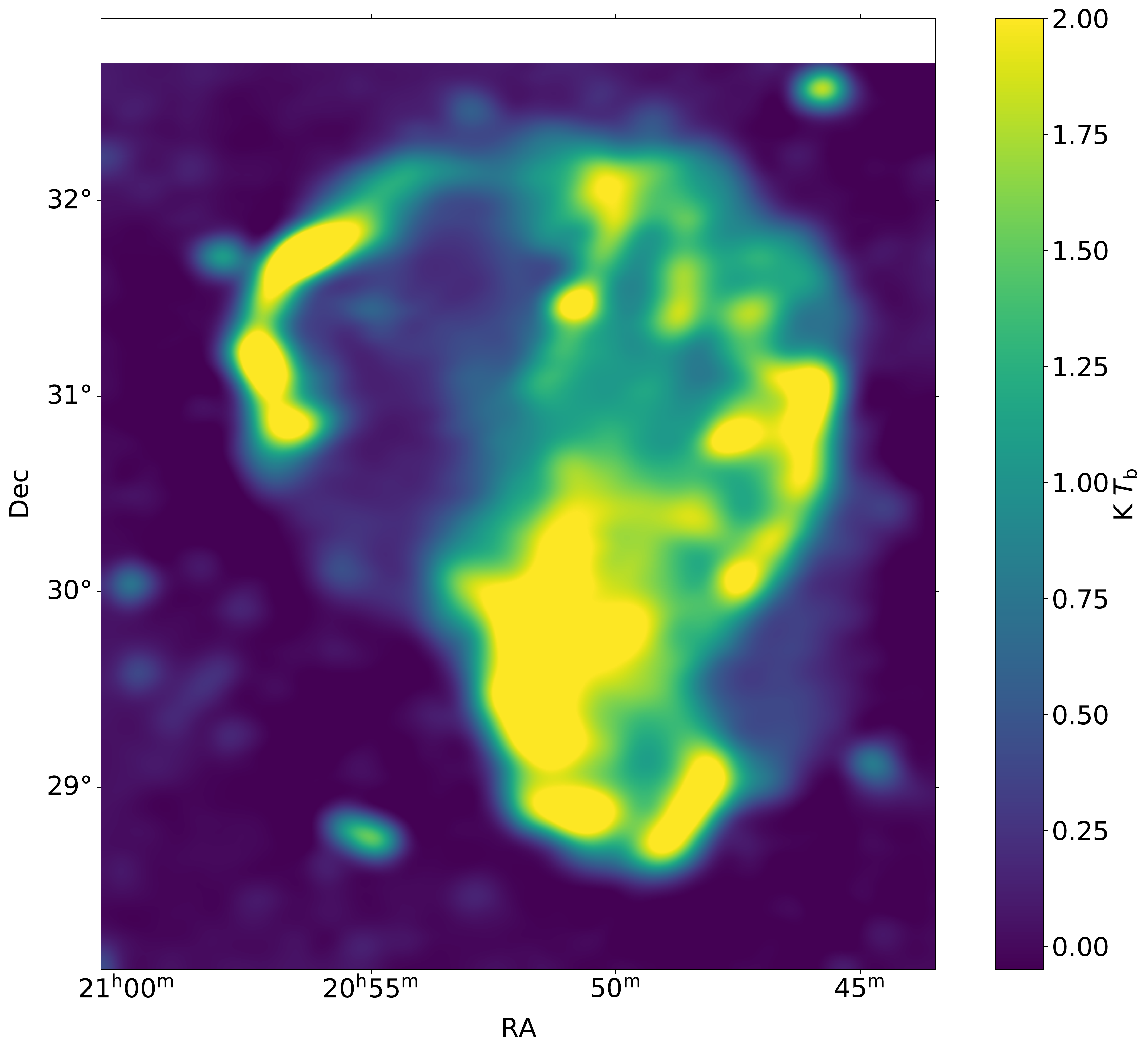}
    \caption{Total intensity images of the Cygnus Loop from FAST at $4\arcmin$ resolution (top panel) and Effelsberg 100-m telescope at $9\farcm4$ resolution (bottom panel).}
    \label{fig:cyg_i}
\end{figure}

We averaged the frequency cube of $I$ by taking the median instead of the mean values to reduce the influence of bad scans caused by RFIs and complex baselines. The corresponding frequency for the averaged map is about 1.28~GHz. At this frequency and the angular resolution of $4\arcmin$, the conversion factor from flux density to brightness temperature is about 12.67~K~$T_{\rm b}$~Jy$^{-1}$ according to Eq.~\ref{eq:jy_k} in Sect.~\ref{sect:gain}. The resulting total-intensity map is shown in Fig.~\ref{fig:cyg_i}. Some scanning effects still remain because the observation was conducted in RA direction only and the $1/f$ noise from low level gain variations and system instabilities thus cannot be completely suppressed. The rms noise measured from the map is about 20~mK~$T_{\rm b}$. The confusion limit at 4$\arcmin$ resolution can be estimated following \citet{condon+74} and \citet{meyers+17} as 
\begin{equation}\label{eq:confusion}
\sigma_c\approx0.2\,\,{\rm mJy\,\, beam^{-1}}\,\,(\frac{\nu}{\rm GHz})^{-0.7}(\frac{\Theta}{\rm arcmin})^2,
\end{equation}
which yields a confusion noise of about 2.7 mJy~beam$^{-1}$, corresponding to about 34~mK~$T_{\rm b}$, larger than the measured rms noise. This could indicate that the confusion limit has been overestimated. \citet{uyaniker+99} measured an rms noise of 15~mK or 7~mJy~beam$^{-1}$ from the Effelsberg Medium Latitude Survey (EMLS), which is below the confusion limit calculated from Eq.~\ref{eq:confusion}. Assuming that the EMLS has reached the confusion limit, the coefficient 0.2 in Eq.~\ref{eq:confusion} should be 0.1 given the frequency of 1.4~GHz and the beam width of $9\farcm4$ for the EMLS. This corresponds to a confusion limit of about 1.35~mJy~beam$^{-1}$ or about 17~mK at 1.28~GHz, close to the measured rms noise from the FAST map. This indicates that the total-intensity observation by FAST has reached the confusion limit. 

For comparison, the total-intensity map from the EMLS~\citep{uyaniker+99} at 1.4~GHz conducted by the 100-m telescope is also shown in Fig.~\ref{fig:cyg_i}. The resolution is about $9\farcm4$. Both the FAST and the EMLS maps clearly exhibit two shell structures of the Cygnus Loop. The FAST map clearly reveals more details of the SNR because of higher angular resolution. For example, the filament inside the northern part, designated as the central filament hereafter, is more outstanding in the FAST map. There are also more extra-galactic sources detected in the FAST map. 

\begin{figure}
    \centering
    \includegraphics[width=0.8\textwidth]{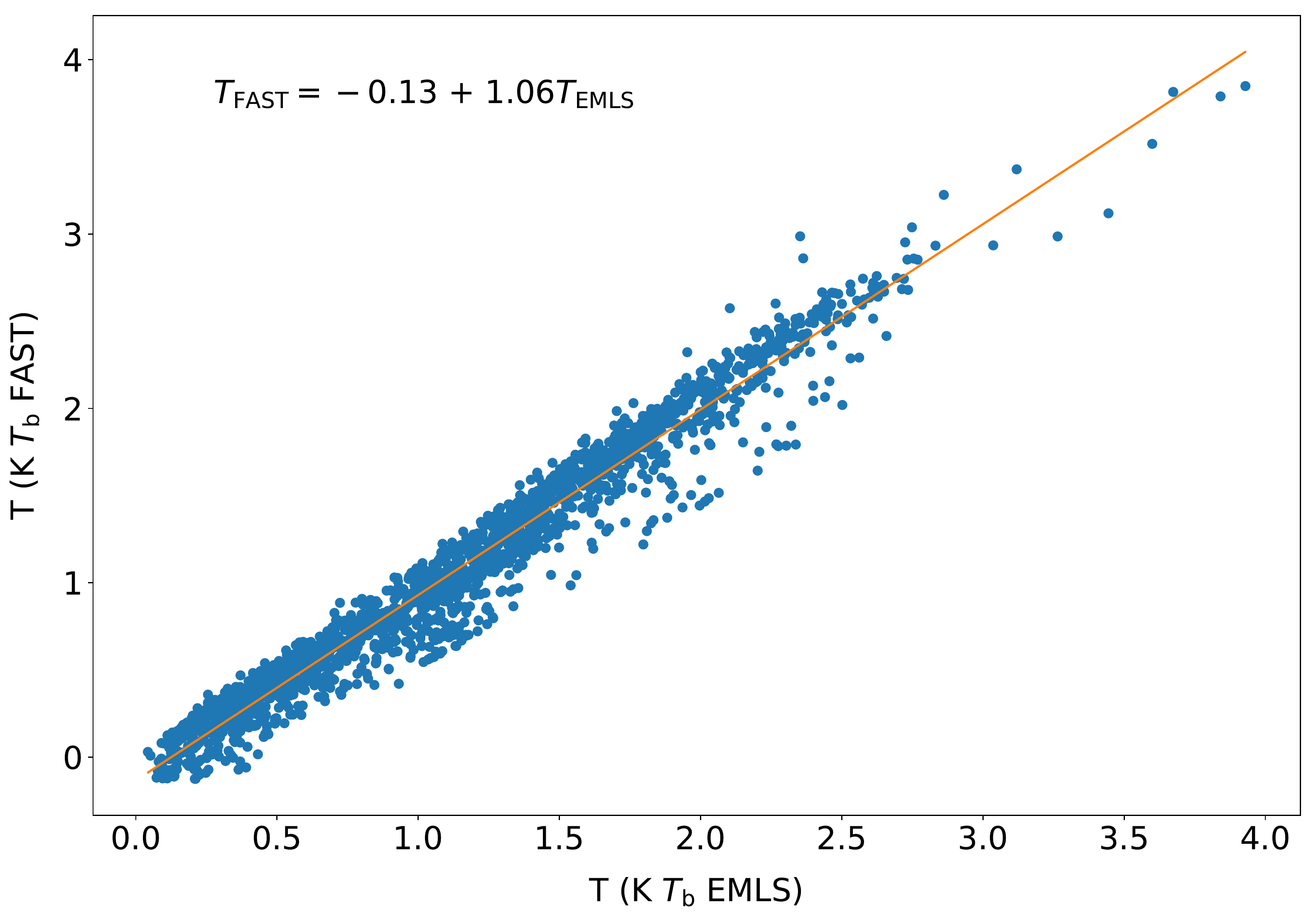}
    \caption{TT-plot between the FAST and EMLS maps for the Cygnus Loop region.}
    \label{fig:tt_fast_emls}
\end{figure}

The temperature versus temperature plot (TT-plot) is a reliable way to determine the relative scale of two maps and hence estimate the brightness temperature spectral index $\beta$, defined as $T_\nu\propto\nu^\beta$, where $\nu$ is the frequency. The flux density ($S$) spectral index $\alpha$, defined as $S_\nu\propto\nu^\alpha$, can be related to $\beta$ as $\beta=\alpha+2$. We used the FAST 1.4~GHz frequency channel map, and smoothed it to the same resolution as the EMLS map. The TT-plot between these two maps for the Cygnus Loop region is shown in Fig.~\ref{fig:tt_fast_emls}. There is a close correspondence of brightness temperature between these two maps, which can be fitted to a straight line as $T_{\rm FAST}=-0.13+1.06\,T_{\rm EMLS}$. This indicates that the map-making procedure for the Cygnus Loop is valid. The FAST temperature is higher than the EMLS by about 6\%, which is very satisfying given that these two maps were derived from different observations with different calibration procedures. The temperature scale of the EMLS data has a scale uncertainty less than about 3\%~\citep{uyaniker+fuerst+98}. The FAST temperature scale has an uncertainty of about 4\% including about 2\% introduced by main beam efficiency fluctuations and about 2\% caused by temperature fluctuations of injected reference signal \citep{jiang+20}. Taking these factors into account, we can conclude that the FAST map agrees well with the EMLS map.

\begin{figure}
    \centering
    \includegraphics[width=0.495\textwidth]{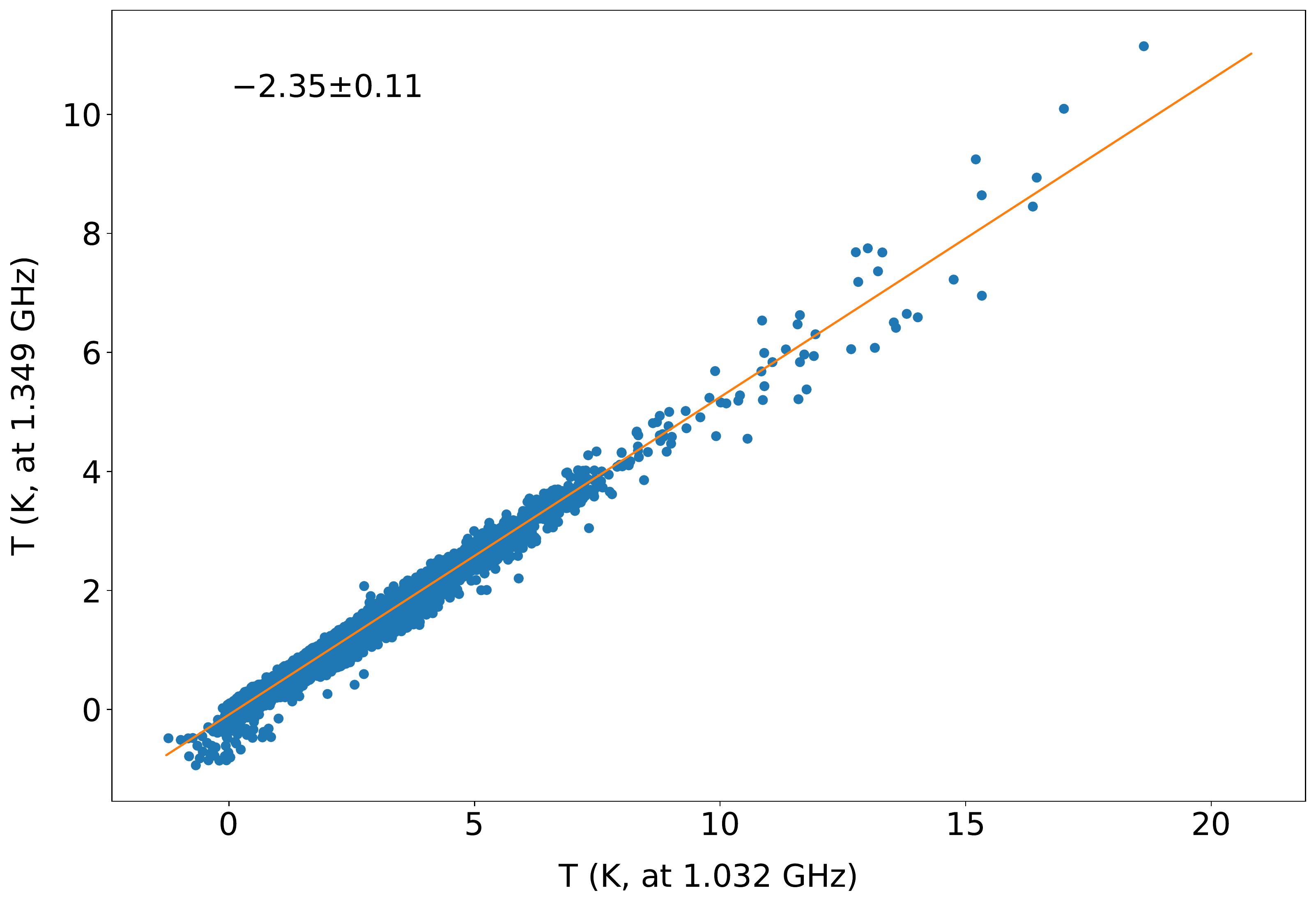}
    \includegraphics[width=0.495\textwidth]{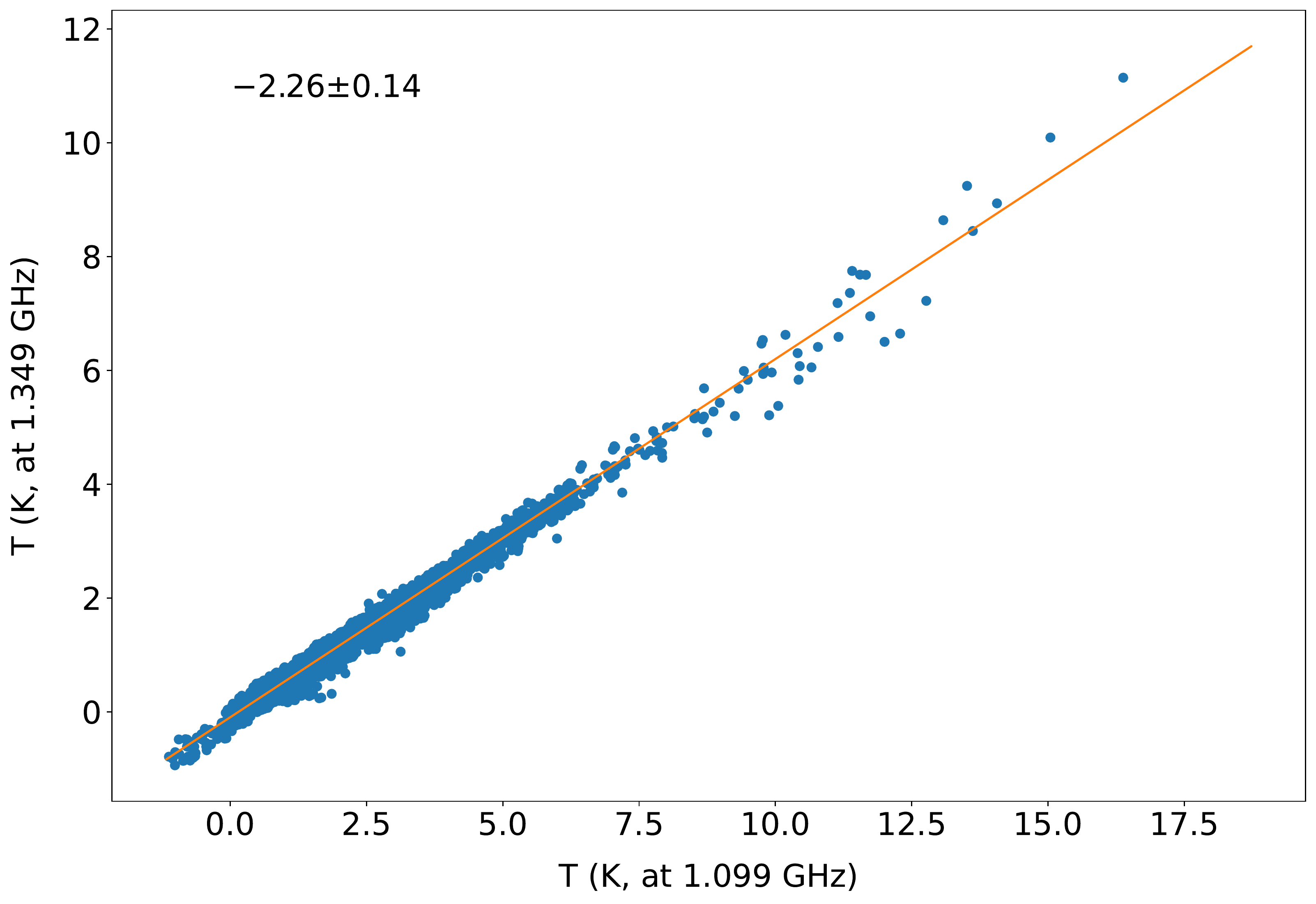}
    \includegraphics[width=0.495\textwidth]{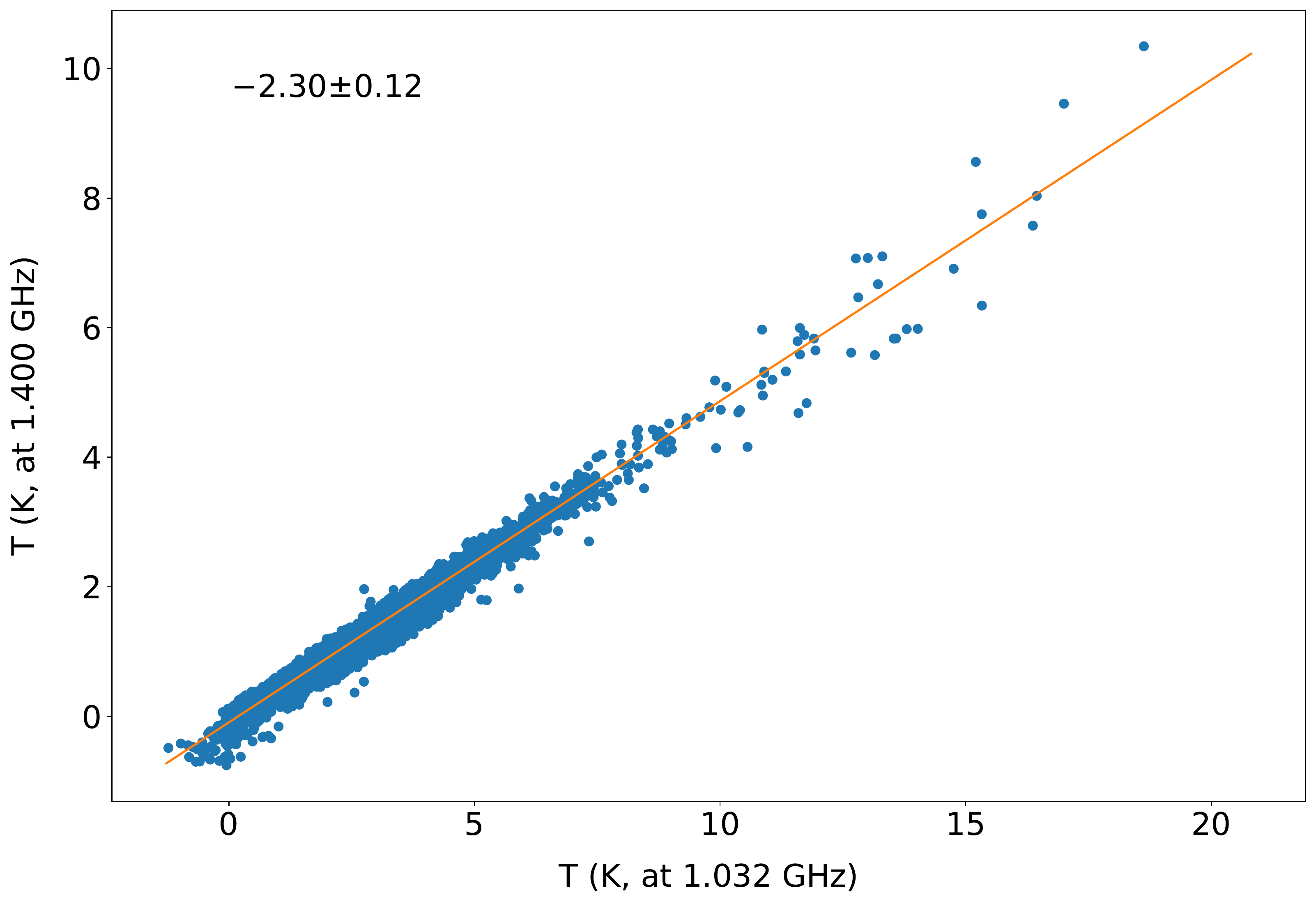}
    \includegraphics[width=0.495\textwidth]{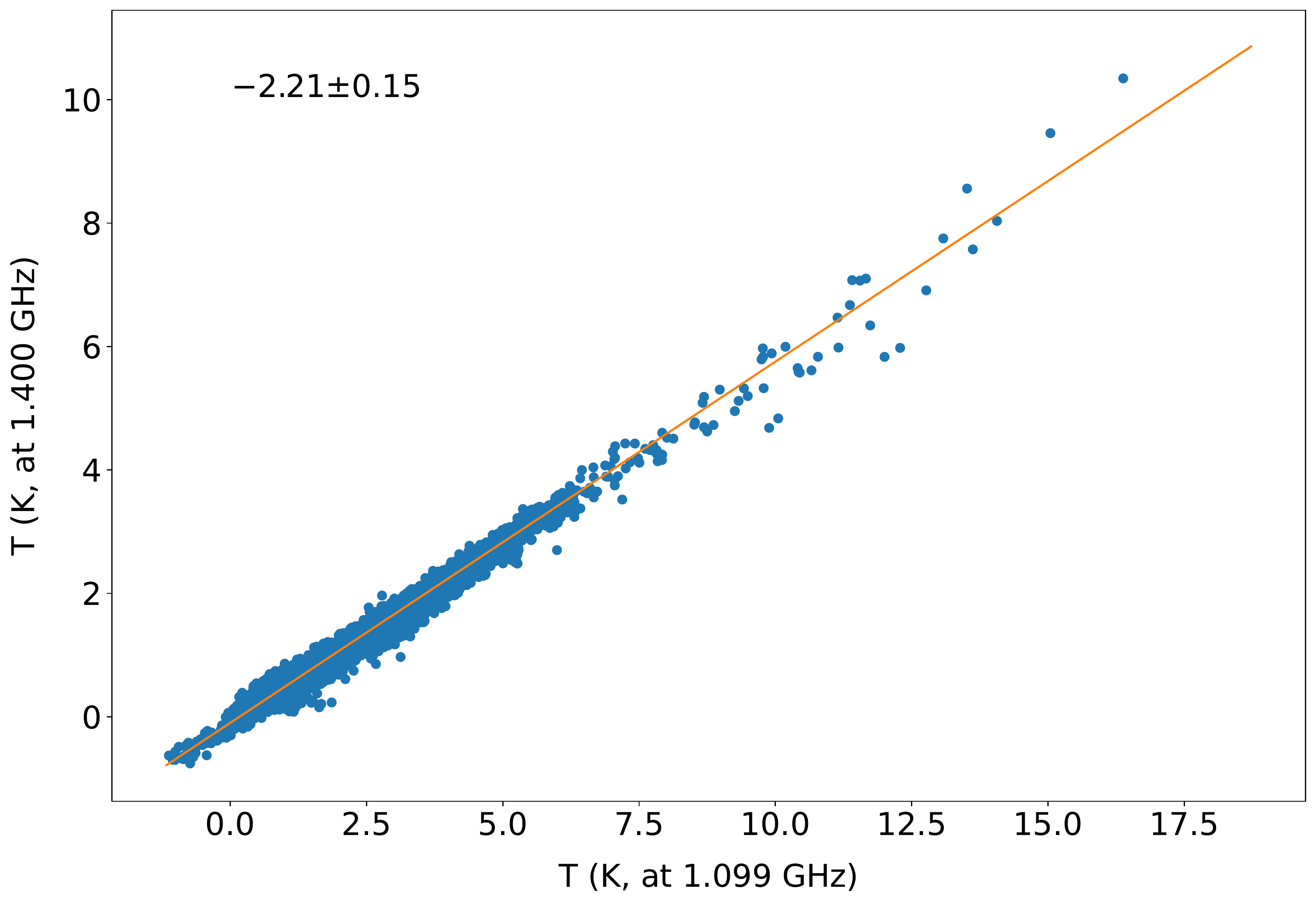}
    \includegraphics[width=0.495\textwidth]{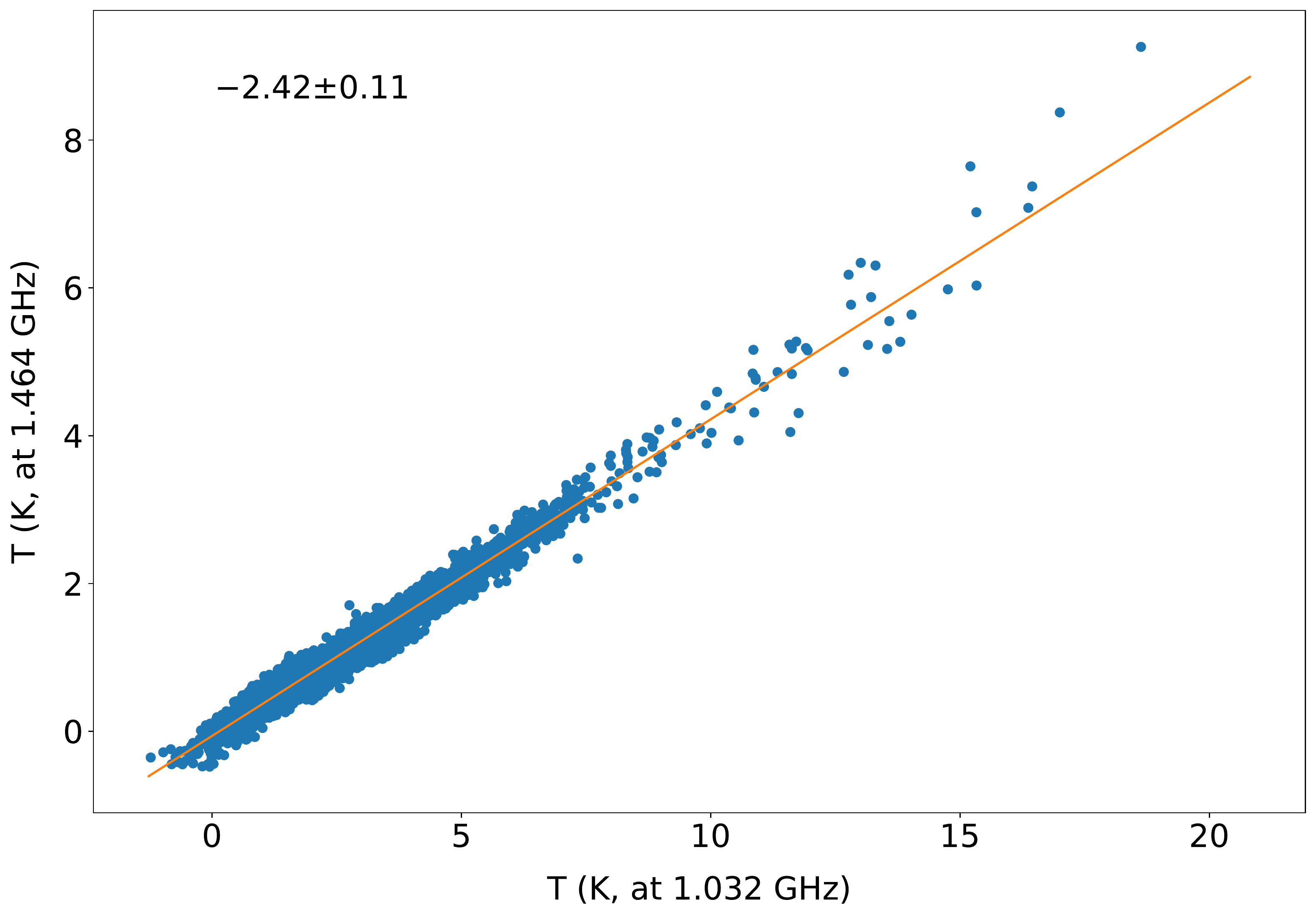}
    \includegraphics[width=0.495\textwidth]{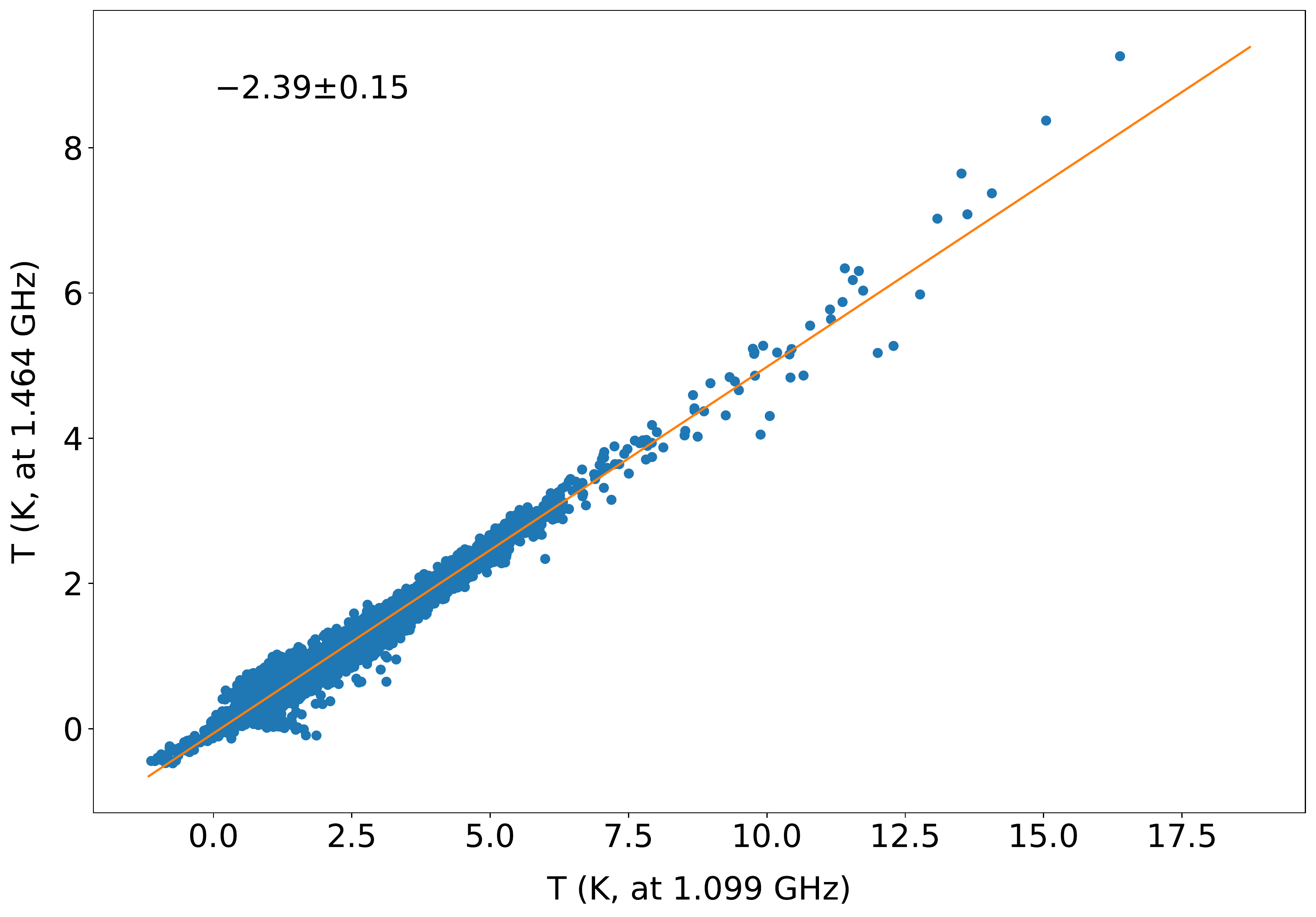}
    \caption{TT-plots between different frequency channels.}
    \label{fig:tt_fast}
\end{figure}

The scales among the FAST frequency channel maps can also be examined. We selected two maps at 1.032 GHz and 1.099 GHz at the low-frequency end, and three frequency channel maps at the middle and high end of the band. The TT-plot between them for the Cygnus Loop region is shown in Fig.~\ref{fig:tt_fast}. The tight correlations between temperatures at different frequency channels again validate the map-making procedure. The spectral index is $\beta=-2.42\pm0.11$ or $\alpha=-0.42\pm0.11$ between the 1.032-GHz and 1.456-GHz channel maps, and $\beta=-2.39\pm0.15$ or $\alpha=-0.39\pm0.15$ between the 1.099-GHz and 1.465-GHz maps. The first pair has nearly the maximal frequency separation of about 430~MHz over the 500-MHz band, and the second pair has a frequency separation of about 370~MHz. These large frequency separations ensure an accurate determination of the spectral index. The spectral indices obtained from both the frequency pairs are consistent with those determined by \citet{sun+06} and \citet{uyaniker+04}, indicating that the scales across the band are appropriate. For small frequency separations, the spectral index from TT-plots is susceptible to brightness temperature fluctuations and more uncertain, and can deviate from the expected values, as can be seen from Fig.~\ref{fig:tt_fast}.

\begin{figure}
    \centering
    \includegraphics[width=0.8\textwidth]{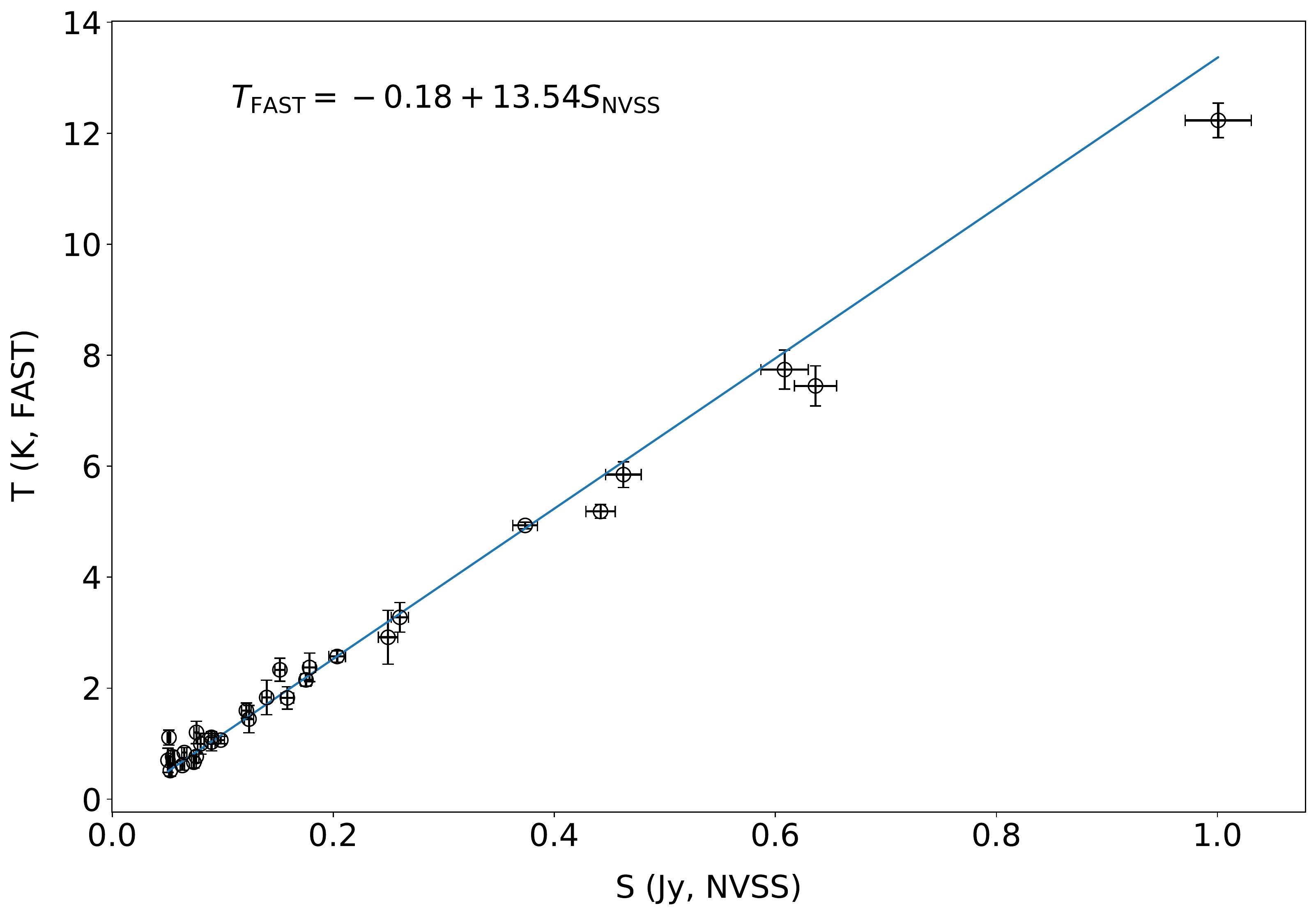}
    \caption{Brightness temperature from FAST versus integrated flux density from NVSS sources.}
    \label{fig:t_s}
\end{figure}

We performed source finding outside the Cygnus Loop area with the Aegean software~\citep{hancock+18} in the FAST total-intensity map in Fig.~\ref{fig:cyg_i} and found 55 compact sources with matches from the NVSS survey~\citep{condon+98}. The median of the position offsets between the FAST and the NVSS sources is about 18$\arcsec$, below 10\% of the $4\arcmin$ beam width. The peak temperature from the FAST map versus the integrated flux density from NVSS is shown in Fig.~\ref{fig:t_s}. Their relation can be fitted by a straight line, which yields a conversion factor of 13.54 K Jy$^{-1}$ from the flux density at 1.4~GHz to the brightness temperature at about 1.28~GHz. For several strong sources with high brightness temperature, there are multiple NVSS sources within the FAST beam, which renders the flux density uncertain. Assuming a spectral index of $-0.8$ for the extra-galactic radio sources~\citep{gasperin+18}, we can extrapolate the flux density from 1.4~GHz to 1.28~GHz, and obtain a conversion factor of about 12.60 K~Jy$^{-1}$ at 1.28~GHz, consistent with the factor of 12.67 K~Jy$^{-1}$ derived from the observations of 3C~138. This means that the flux-density scale can be determined from the NVSS sources instead of observations of the calibrators, which will be essential for the CRAFTS~\citep{li+18} survey by FAST.

The integrated flux density of the Cygnus Loop is about 164 Jy with the conversion factor of 12.67 K~Jy$^{-1}$. In contrast, the flux density at 1420~MHz is $143\pm14$~Jy measured by \citet{uyaniker+04}, corresponding to a flux density of about 149~Jy at 1.28~GHz assuming a spectral index $\alpha=-0.42$. The integrated flux density by FAST is about 10\% larger than that by EMLS, which is expected as the FAST brightness temperature is about 6\% higher than the EMLS brightness temperature (Fig.~\ref{fig:tt_fast_emls}). Note that the main calibrator for the EMLS is 3C286, which is 
stable, while the flux density of 3C~138 is highly variable~\citep{perley+butler+13}, which could also contribute to the difference of the flux densities. Based on the model by \citet{perley+butler+17}, the flux density of 3C~138 is about 8.99~Jy at 1.28~GHz and about 8.56~Jy at 1.4~GHz.  

\subsection{Polarization}

The observed polarization intensity $P(\lambda^2)$ can be written as \citep{burn+66},
\begin{equation*}
    P(\lambda^2)=Q(\lambda^2)+jU(\lambda^2)=\int_{-\infty}^{+\infty} F(\phi) \exp{(2j\phi\lambda^2)}{\rm d}\phi
\end{equation*}
where $\phi(\vec{r})$ is the Faraday depth defined as the integral of the thermal electron density weighted by the line-of-sight magnetic field from the position $\vec{r}$ inside the source to the observer, and $F(\phi)$ is the Faraday spectrum or the Faraday dispersion function, representing the polarization intensity as a function of Faraday depth. For a source within only one single emitting component and without internal Faraday rotation, $F(\phi)$ is a $\delta$ function and the peak $\phi$ is equivalent to the RM.

There are many algorithms available to reconstruct $F(\phi)$ from $P(\lambda^2)$~\citep{sun+15}, among which RM synthesis \citep{brentjens+05} is widely used. The method is based on that $P(\lambda^2)$ and $F(\phi)$ are Fourier transform pairs. With RM synthesis, the Faraday spectrum can be derived as,
\begin{align*}
    \widetilde{F(\phi)} &= \frac{\sum W(\lambda_i^2)P(\lambda_i^2)\exp{[-2j\phi(\lambda_i^2-\lambda_0^2)]}}{\sum W(\lambda_i^2)}=F(\phi) * R(\phi)\\
    \lambda_0^2 & = \frac{\sum W(\lambda_i^2)\lambda_i^2}{\sum W(\lambda_i^2)}\\
    R(\phi) & = \frac{\sum W(\lambda_i^2)\exp{[-2j\phi(\lambda_i^2-\lambda_0^2)]}}{\sum W(\lambda_i^2)}
\end{align*}
where $\lambda_i$ is the wavelength of the frequency channel, $W(\lambda_i^2)$ is the weight for each individual channel, the RM spread function (RMSF) $R(\phi)$ is the window function determined by the sampling in $\lambda^2$ domain, and the sum is over all the frequency channels. Because of a finite bandwidth, RM synthesis yields a Faraday spectrum which is the convolution of true $F(\phi)$ with $R(\phi)$. A de-convolution can be performed on $\widetilde{F(\phi)}$ to obtain $F(\phi)$ by employing RM clean~\citep{heald+09}. The width of the RMSF is about 90~rad~m$^{-2}$ for the frequency range of FAST. 

\begin{figure}
    \centering
    \includegraphics[width=0.8\textwidth]{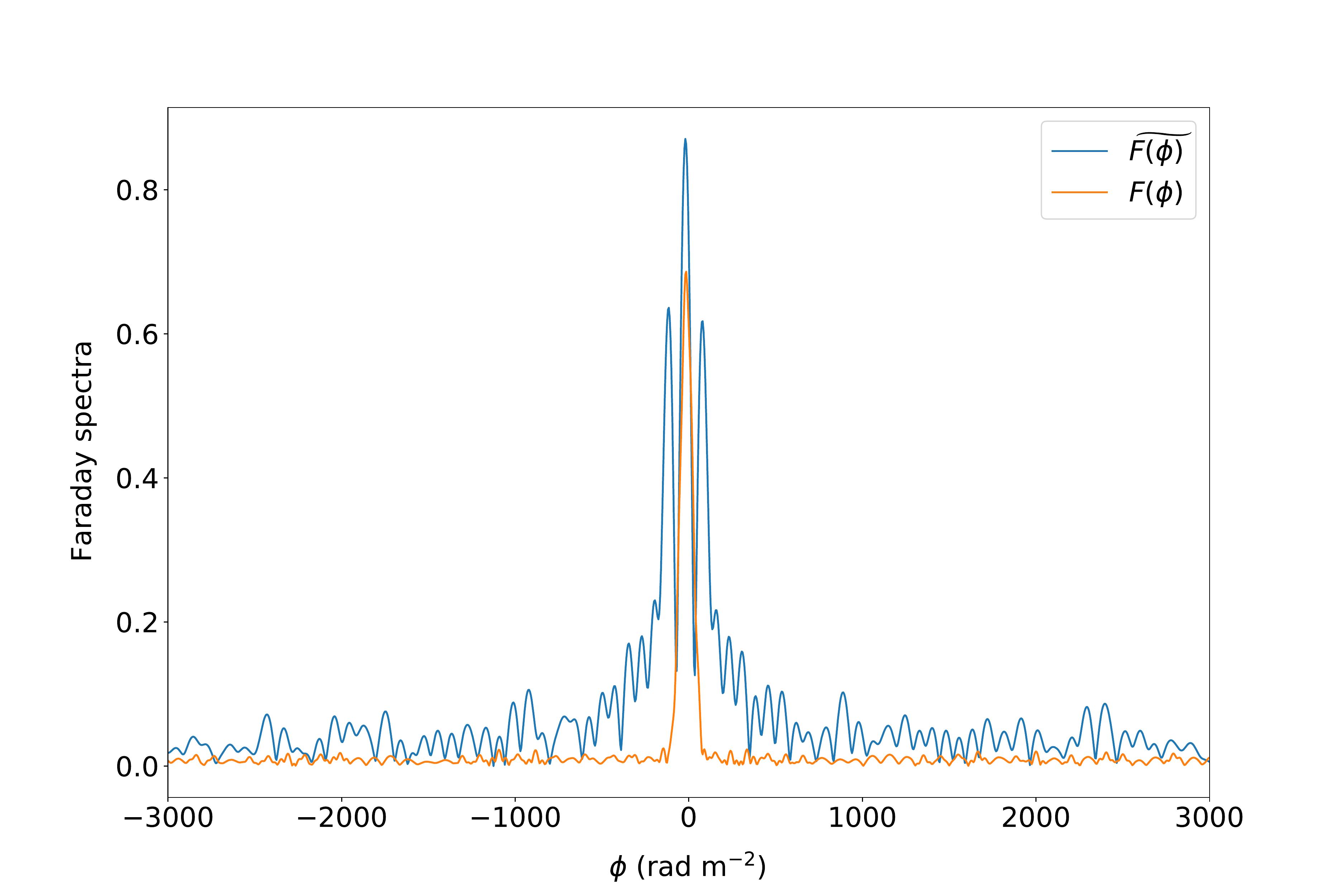}
    \caption{Faraday spectra before (blue) and after RM clean (red).}
    \label{fig:rms_prof}
\end{figure}

We applied RM synthesis to the frequency cubes of $Q$ and $U$ of the Cygnus Loop and obtained the Faraday depth cubes of $\widetilde{F(\phi)}$. The Faraday depth is set to be $-10^4 \leq \phi \leq10^4$~rad~m$^{-2}$ in a step interval of 5~rad~m$^{-2}$. We then applied RM clean to derive the true $F(\phi)$. An example of the Faraday spectrum before and after RM clean is shown in Fig.~\ref{fig:rms_prof}. Because of frequency gaps caused by RFI, strong sidelobes are present in the original Faraday spectrum from RM synthesis and are removed with RM clean.

\begin{figure}
    \centering
    \includegraphics[width=0.8\textwidth]{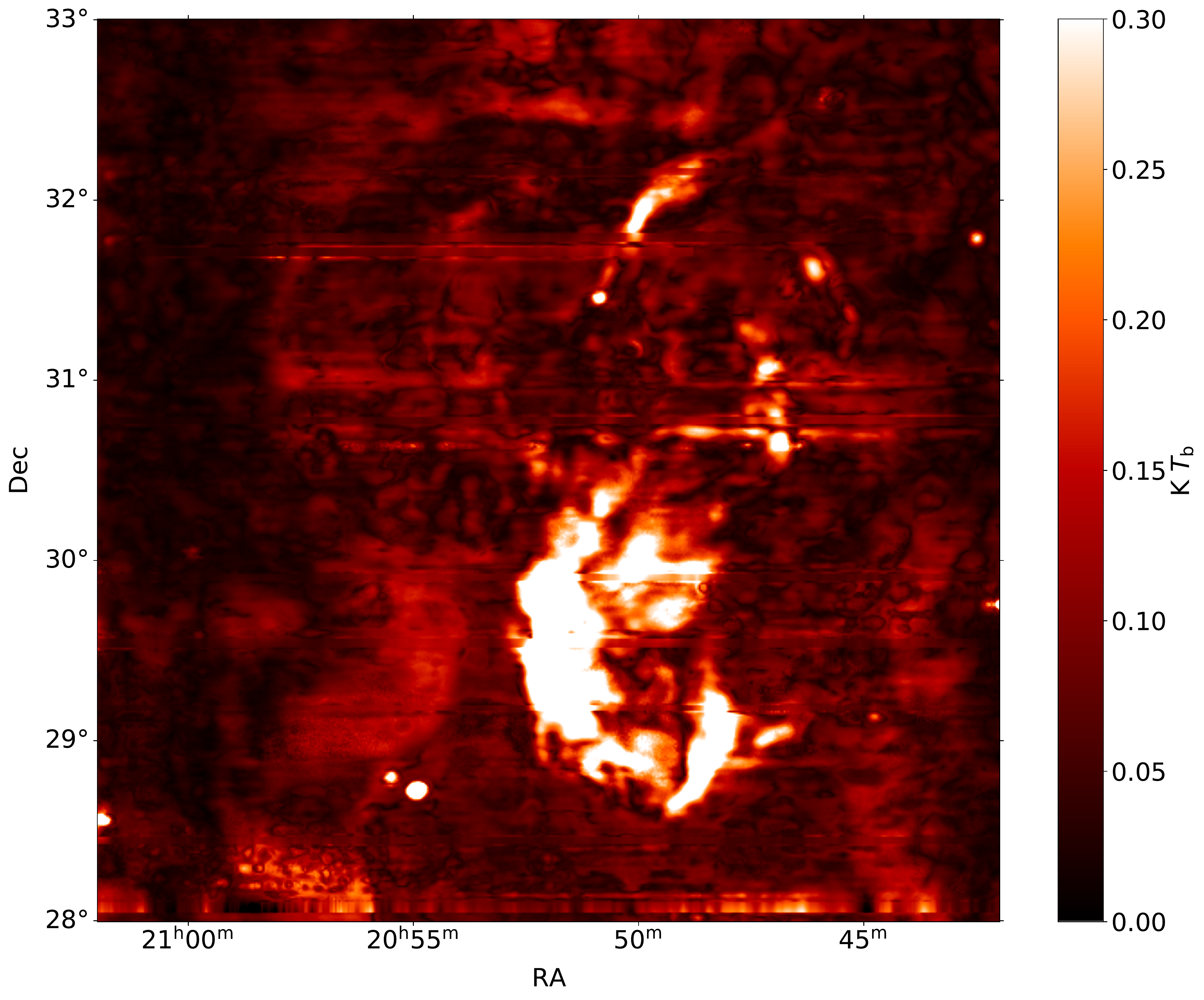}
    \includegraphics[width=0.8\textwidth]{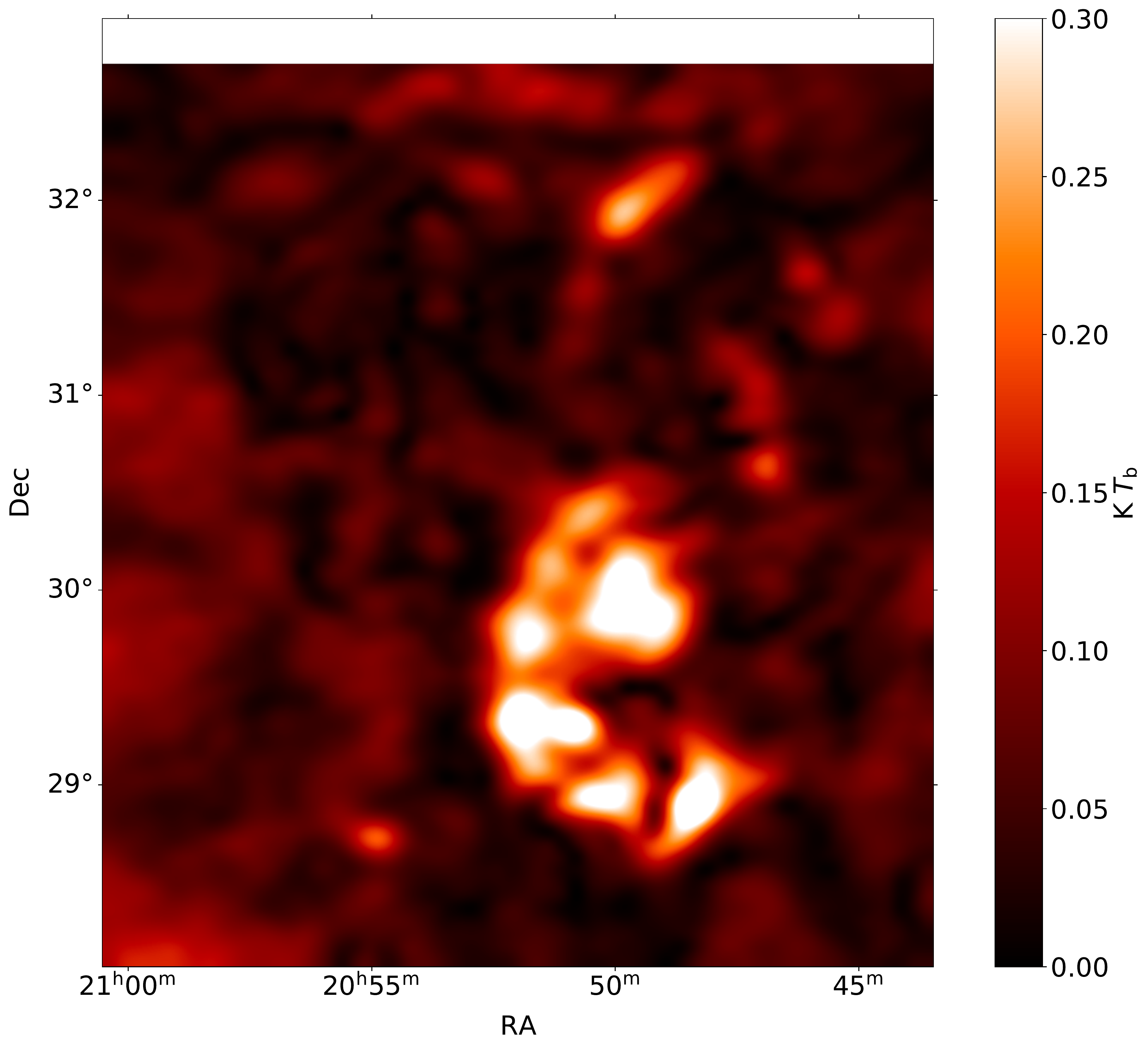}
    \caption{Polarized-intensity images of the Cygnus Loop from FAST from RM-synthesis (top panel) and Effelsberg 100-m telescope (bottom panel).}
    \label{fig:cyg_pi}
\end{figure}

We searched for the peaks of $|F(\phi)|$ and obtained the polarization-intensity map shown in Fig.~\ref{fig:cyg_pi}. For comparison, we also display the polarization-intensity map from EMLS. The polarized emission from the southern shell, the central filament and some segments of the northern shell can be clearly seen in the FAST map, which are much sharper than in the EMLS map. Several polarized sources are also detected in the FAST map. The rms noise is about 5~mK~$T_{\rm b}$. We also averaged $Q$ and $U$ over the full FAST band by taking medians of all the frequency channels, and then calculated polarized intensity. The rms noise measured from the averaged $Q$, $U$, and polarized intensity are all about 5~mK~$T_{\rm b}$. The rms noise in polarized intensity is much lower than in the total-intensity map, implying that the confusion limit is much lower in polarization and the rms noise of the total-intensity has probably reached the level of confusion limit.

\begin{figure}
    \centering
    \includegraphics[width=0.8\textwidth]{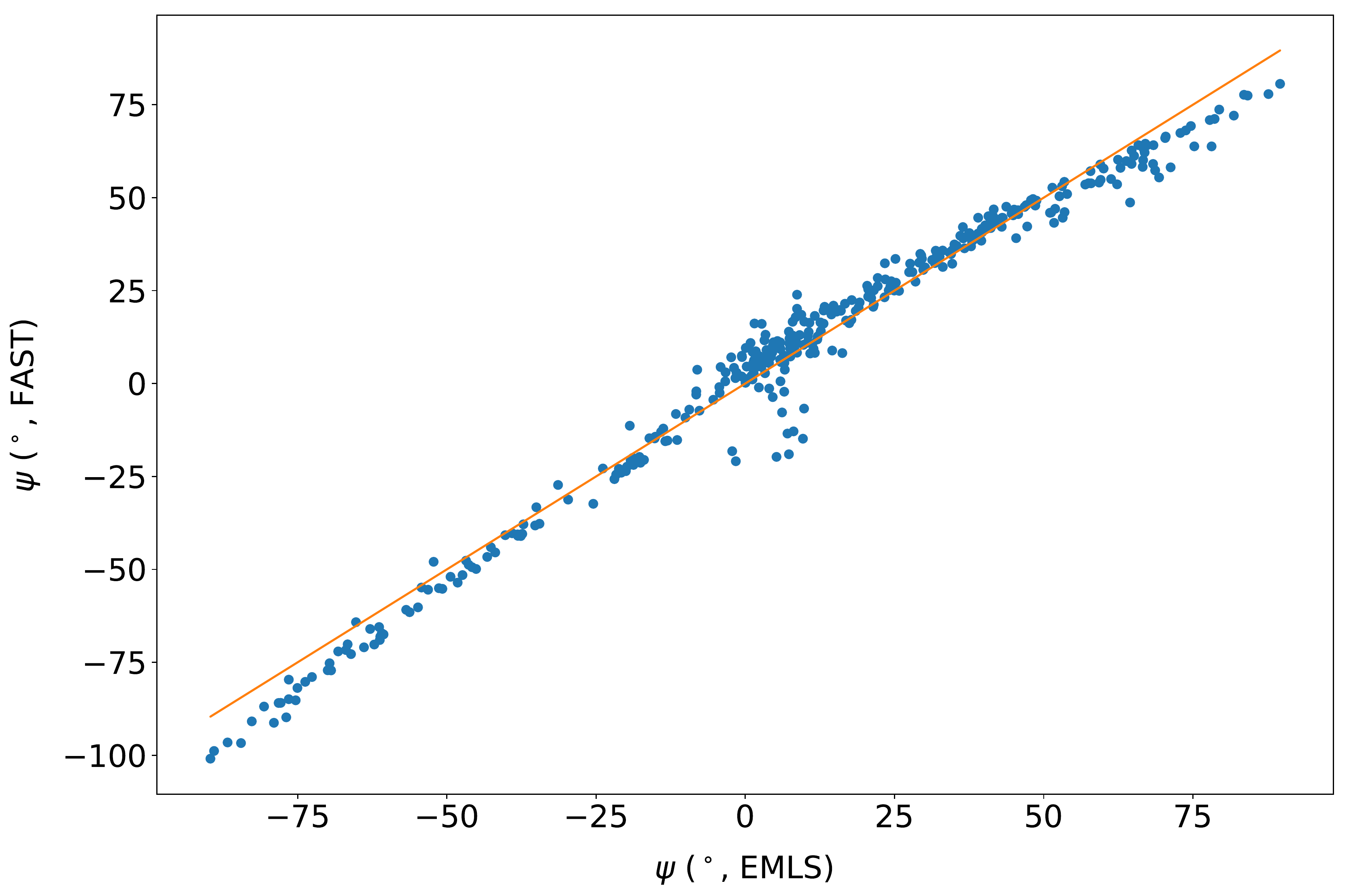}
    \caption{Polarization angle from FAST versus EMLS for the southern shell of the Cygnus Loop with the line of equality.}
    \label{fig:pa_fast_emls}
\end{figure}

\begin{figure}
    \centering
    \includegraphics[width=0.8\textwidth]{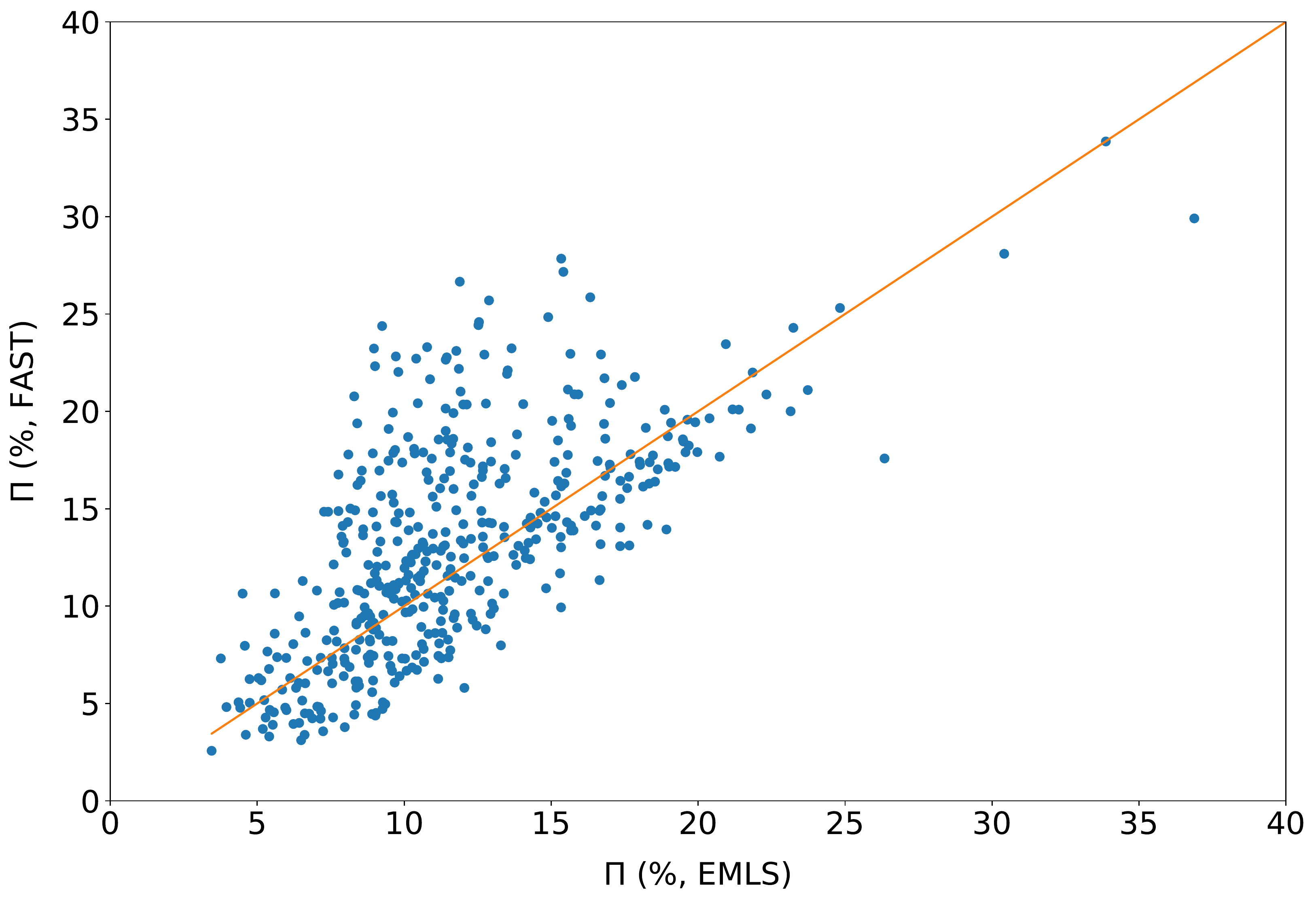}
    \caption{Polarization percentage from FAST versus EMLS for the southern shell of the Cygnus Loop with the line of equality.}
    \label{fig:pi_fast_emls}
\end{figure}

We selected $I$, $Q$ and $U$ maps at 1.4~GHz from the FAST frequency cubes of the Cygnus Loop, smoothed them to the same resolution as the EMLS map, and calculated the polarization angle $\psi$ and the polarization percentage $\Pi$. According to the TT-plot (Fig.~\ref{fig:tt_fast_emls}), we added 130~mK to the FAST total intensity map to adjust the background level when calculating the polarization fraction. We show $\psi$ and $\Pi$ of the bright southern shell~(Fig.~\ref{fig:cyg_pi}) from FAST against those from the EMLS in Fig.~\ref{fig:pa_fast_emls} and Fig.~\ref{fig:pi_fast_emls}, respectively. For polarization angles, the values from FAST agree with those from EMLS. The difference has a median of about $-0\fdg6$, and a scattering of about 3$\degree$. When the polarization angle approaches $0\degree$ or $90\degree$, the difference is more pronounced. This could be caused by instrumental polarization that has not yet been corrected. For polarization percentage, the FAST values are larger than the EMLS values in the range of 10\%-15\%, implying that depolarization occurs in the EMLS map. The depolarization is probably caused by the turbulence of the thermal gas in front of the emitting region, which is called beam depolarization~\citep{sokoloff+98}. The EMLS has a large beam width and therefore experiences more beam depolarization. The beam depolarization does not influence the polarization angle as seen from Fig.~\ref{fig:pa_fast_emls}.

\begin{figure}
    \centering
    \includegraphics[width=0.8\textwidth]{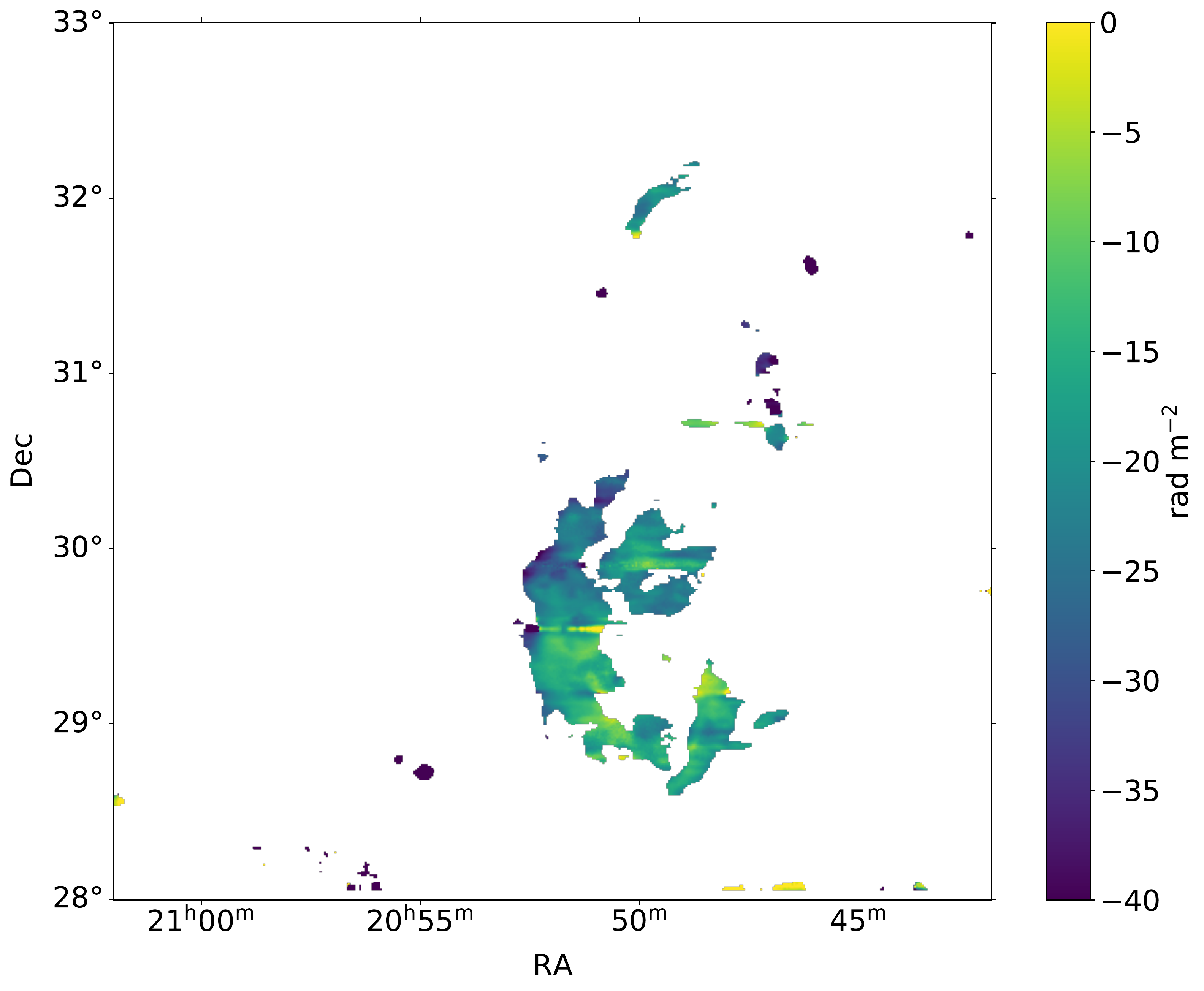}
    \caption{RM map of the Cygnus Loop at 4\arcmin angular resolution.}
    \label{fig:rm}
\end{figure}

Most of the Faraday spectra have only one peak after RM clean, as shown in Fig.~\ref{fig:rms_prof}. Therefore, the peak Faraday depth is approximately equivalent to the RM. We made parabola fit to $|F(\phi)|$ near the peaks and obtain peak $\phi$ for the southern shell and the central filament of the Cygnus Loop. The resulting map is shown in Fig.~\ref{fig:rm}. For the southern shell, the mean RM is about $-20$~rad~m$^{-2}$, consistent with the value of $-$21~rad~m$^{-2}$ derived by \citet{sun+06} based on data at 4800~MHz and 2695~MHz. For the central filament, the average RM is about $-$19~rad~m$^{-2}$, very similar to that of the southern shell. 

\begin{figure}
    \centering
    \includegraphics[width=0.8\textwidth]{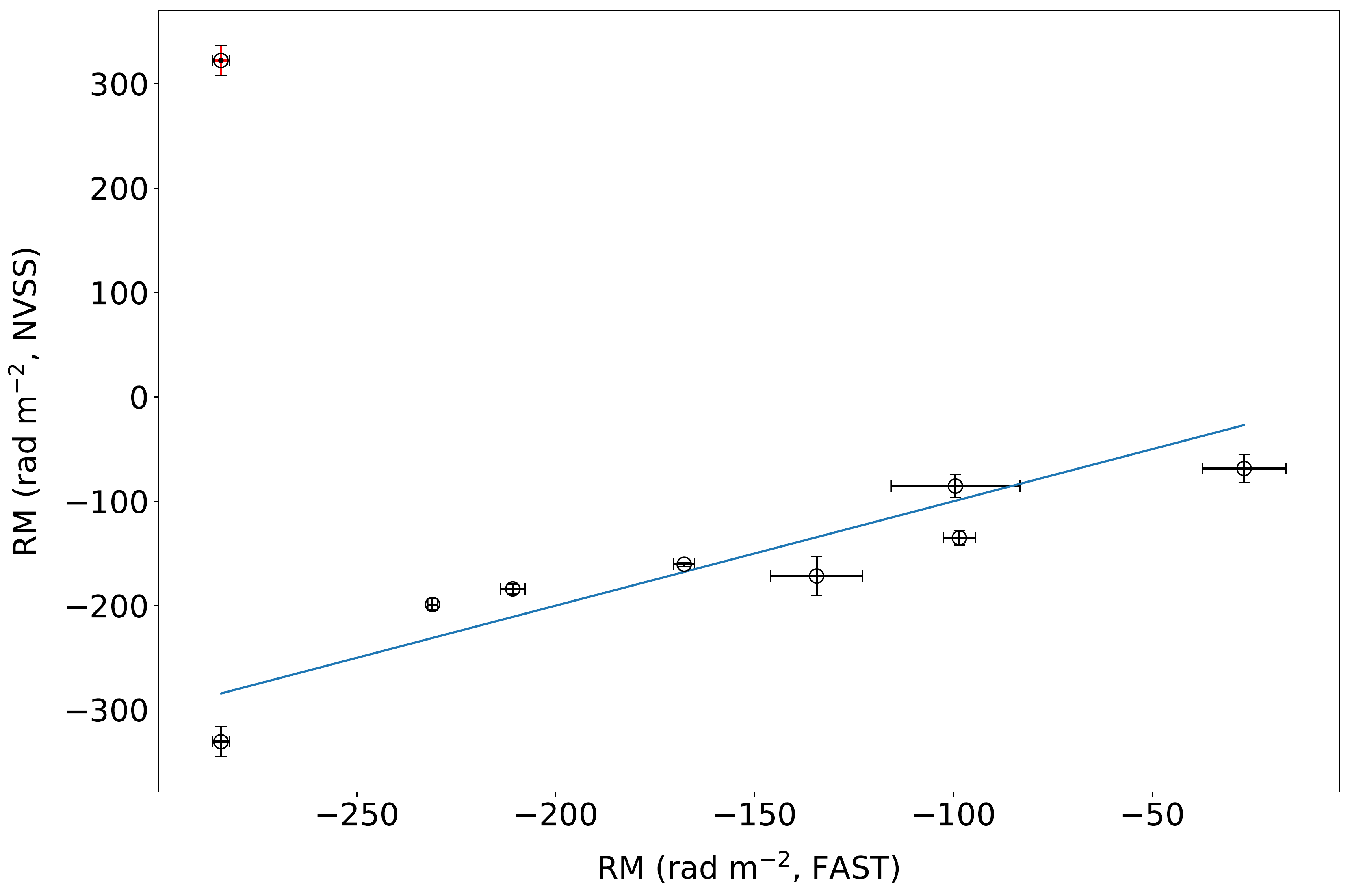}
    \caption{RMs from FAST versus NVSS with the line of equality.}
    \label{fig:rm_source}
\end{figure}

We detected eight polarized sources, and obtained their RMs. We also retrieved the RMs from the NVSS RM catalogue by \citet{taylor+09}. The RMs are shown in Fig.~\ref{fig:rm_source}. For the source at ${\rm (RA,\, Dec)}=(20^{\rm h}50^{\rm m}52\fs36,\, +31\degr27\arcmin31\farcs49)$, the RM from NVSS is 322.5~rad~m$^{-2}$, whereas the RM from FAST is $-$284.2~rad~m$^{-2}$. Because the NVSS RMs are calculated based on two closely separated frequency bands and there are $n\pi$-ambiguities for polarization angles, the RMs have ambiguity of multiples of 652.9~rad~m$^{-2}$. After subtraction of 652.9~rad~m$^{-2}$, the NVSS RM is consistent with the FAST RM. This demonstrates that FAST is able to overcome the ambiguity of RMs with the wide band and multi-frequency channels. Actually, the RM of this source is an outlier as all of the neighbouring RMs are negative. We followed \citet{ma+19} to calculate the median ($\rm RM_{3\degree}$) and standard deviation ($\sigma_{3\degree}$) of RMs within a $3\degree$-radius of the source, and then derived $\Delta/\sigma \equiv |\rm RM - RM_{3\degree}|/\sigma_{3\degree}\approx5.3$. This value satisfies the criterion value of about 2.85 that \citet{ma+19} obtained for an ambiguous RM. Note that \citet{ma+19} studied RM ambiguities outside the Galactic plane with Galactic latitudes $|b|>10\degree$. Towards the Galactic plane $|b|<10\degree$, the RM contribution from the Galaxy increases significantly, more sources with large $\rm |RM|$ values could be influenced by ambiguities.    

\section{Conclusions}
\label{sect:conclusion}

We present continuum and polarization observations of the Cygnus Loop by FAST, and described the calibration and map-making procedures. By comparing the brightness temperature, the in-band spectral index, the polarization angle, and RM of the FAST measurements with previous measurements, we verified the data processing for continuum and polarization imaging. Particularly, the polarization has a sensitivity of about 5~mK and seems not to be affected by the confusion limit, which provides a unique opportunity for polarization observations of faint diffuse structures. We demonstrate that FAST can determine RMs from compact sources more accurately than NVSS by overcoming the RM ambiguities with multi-frequency channel wide band.  

\begin{acknowledgements}
XS and XL are supported by the National Natural Science Foundation of China (NSFC) under No.11763008, and the Science \& Technology Department of Yunnan Province - Yunnan University Joint Funding (2019FY003005). XS and XG are supported by the CAS-NWO cooperation programme (Grant No. GJHZ1865) and XG is supported by the National Natural Science Foundation of China (Grant No. U1831103). The project was supported by the Key Lab of FAST, National Astronomical Observatories, Chinese Academy of Sciences. We thank Dr. Patricia Reich for reading the manuscript and discussions. We also thank Dr. Lei Qian for helping schedule the observations, and Dr. Marko Kr\u{c}o and Dr. Meng-Ting Liu for providing us the measurements of the temperatures of the injected noise. We thank the referee for the comments that have improved the paper.
\end{acknowledgements}

\bibliographystyle{raa}
\bibliography{bibtex}
\end{document}